\documentclass[11pt]{article}
\usepackage{natbib}
\usepackage{tabularx} % extra features for tabular environment
\usepackage{amsmath,amsthm,amssymb}
\usepackage{bm,bbm}% improve math presentation
\usepackage{graphicx} % takes care of graphic including machinery
\usepackage[margin=1in]{geometry} % decreases margins
\usepackage{float}
\usepackage{multirow}
\usepackage{setspace}
\usepackage[final]{hyperref} % adds hyper links inside the generated pdf file
\hypersetup{
	colorlinks=true,       % false: boxed links; true: colored links
	linkcolor=blue,        % color of internal links
	citecolor=blue,        % color of links to bibliography
	filecolor=magenta,     % color of file links
	urlcolor=blue         
}
\def\R{\mathbbm{R}}
\def\E{\mathbbm{E}}
\def\P{\mathbbm{P}}
\def\calA{\mathcal{A}}

\def\bx{\bm{x}}
\def\by{\bm{y}}
\def\bg{\bm{g}}
\def\Sig{\Sigma}
\def\sig{\sigma}
\def\gam{\gamma}
\def\Gam{\Gamma}

\def\bbeta{\bm{\beta}}
\def\eps{\epsilon}

\def\bM{\mathcal{M}}
\DeclareMathOperator* {\argmin} {arg\,min}

\usepackage[ruled]{algorithm2e}
\usepackage{algpseudocode}

\newtheorem{theorem}{Theorem}

\newtheorem{lemma}{Lemma}
\newtheorem{remark}{Remark}
\newtheorem{condition}{Condition}
\usepackage{authblk}
\title{\textbf{Multi-dimensional domain generalization with low-rank structures}}
\author[1]{Sai Li}
\author[2]{Linjun Zhang}
\affil[1]{Institute of Statistics and Big Data, Renmin University of China}
\affil[2]{Department of Statistics, Rutgers University}
\date{}
\begin{document}

\maketitle 

\begin{abstract}
In conventional statistical and machine learning methods, it is typically assumed that the test data are identically distributed with the training data. However, this assumption does not always hold, especially in applications where the target population are not well-represented in the training data. This is a notable issue in health-related studies, where specific ethnic populations may be underrepresented, posing a significant challenge for researchers aiming to make statistical inferences about these minority groups. In this work, we present a novel approach to addressing this challenge in linear regression models. We organize the model parameters for all the sub-populations into a  tensor. By studying a structured tensor completion problem, we can achieve robust domain generalization, i.e., learning about sub-populations with limited or no available data. Our method novelly leverages the structure of group labels and it can produce more reliable and interpretable generalization results.  We establish rigorous theoretical guarantees for the proposed method and demonstrate its minimax optimality. To validate the effectiveness of our approach, we conduct extensive numerical experiments and a real data study focused on education level prediction for multiple ethnic groups, comparing our results with those obtained using other existing methods.
%Conventional statistical and machine learning methods always assume the test data are identically distributed as the training data. In many applications, however, the target population may not be well-represented in the training data. In health-related studies, for instance, certain ethnic populations may be underrepresented, which presents a challenge for researchers seeking to statistically learn these minority groups. In this work, we tackle this challenge in linear models by organizing the model parameters of all the sub-populations into a tensor. We formulate the domain generalization problem as a tensor completion task, allowing us to learn about sub-populations with limited or no available data. Our method novelly leverages the structure of group labels which can produce more reliable and interpretable generalization results. We establish the theoretical guarantees for our proposal and further prove its minimax optimality. Numerical experiments and a real data study on salary prediction for ten ethnic groups are conducted with our proposal and other comparable methods. 
\end{abstract}

\section{Introduction}
%Conventional machine learning methods assume that the test data, which are sampled from a learning target, are well-represented in the training domains. However, data from target domains can be limited or even unseen in the training phase. For instance, in biomedical researches, different clinical centers can have different medical devices and patient demography which can lead to the discrepancy between the training and test distributions \citep{zhang2023federated}. Due to multiple levels of heterogeneity in the test and training domains,  the test data can have different distributions from the training data. Domain generalization, which is also known as out-of-distribution generalization or zero-shot learning,  considers the problem of learning a model which can generalize well when the target population is not represented in the training domains \citep{wang2022generalizing,zhou2022domain}. 
Conventional machine learning methods typically assume that the test data, sampled from a target distribution, are well-represented in the training domains. However, in many practical scenarios, data from the target domains can be scarce or completely unseen during the training phase. A prominent example of this occurs in biomedical research: different clinical centers may employ varied medical devices and serve diverse patient demographics, leading to significant discrepancies between the training and test distributions \citep{zhang2023federated}. Such scenarios introduce multiple levels of heterogeneity between the test and training domains, causing the test data distributed differently from the training data. To address this challenge, the field of \textit{domain generalization} has emerged. Also referred to as out-of-distribution generalization or zero-shot domain adaptation, domain generalization focuses on the development of models that can effectively generalize to new and even unseen populations which are not represented in the training domains \citep{wang2022generalizing,zhou2022domain}.

Domain generalization has attracted significant attention from various disciplines, including computer vision, healthcare, and biological studies \citep{hendrycks2021many, lotfollahi2021out, sharifi2021out}. Consider a typical example where the goal is to train a classification model to distinguish between cats and dogs based on images. If the training data consists of images in cartoon or painting styles, while the test data comprises real photos, there is a clear divergence between the style features in the training and test domains. In such cases, a classification rule trained on the artistic images may not perform well when applied to real photos. In health-related studies, this issue becomes even more critical. Populations with varying demographic features, such as gender and race, may exhibit distinct biological mechanisms underlying diseases \citep{woodward2022genetic}. Moreover, certain sub-populations may be underrepresented in medical databases, exacerbating the challenge of ensuring that predictive models--trained on well-represented groups--are generalizable and beneficial to those underrepresented populations. This underscores the urgent need for reliable domain generalization methods. %Pursuing such inclusivity is essential for gaining more comprehensive insights, ensuring equitable healthcare solutions, and developing robust and reliable models for complex real-world applications.

%Domain generalization has received much attention from different fields including computer vision, healthcare, and biological studies \citep{hendrycks2021many,lotfollahi2021out, sharifi2021out}. As a typical example, suppose that we aim at training a classification model to tell whether an animal in an image is a cat or a dog. The training images are in cartoon or painting styles but the test images are real photos. In this case, the style features are different in the training and test domains and the classification rule learned based on the training domain may not perform well in the test domain. In health-related studies, populations with different demographic features (e.g., gender and race) can have distinct mechanisms underlying disease \citep{woodward2022genetic}. As some sub-populations are disproportionately represented in certain databases, there is a pressing need to ensure that models derived from well-represented groups can be generalized to benefit underrepresented groups. This pursuit of inclusivity ensures more comprehensive insights and equitable healthcare solutions, and is more robust and reliable in complicated real applications. 
%In summary, a method with satisfactory domain generalization properties is more robust and reliable in complicated real applications. 

\subsection{Main results}
\label{sec1-1}
In this work, we consider a multi-task linear regression framework. Suppose that we have observations $(\bx_i^\top ,y_i, \bg_i^\top )$, for $i=1,\dots,N$, where $\bx_i\in\R^p$ represents the vector of covariates, $y_i\in\R$ is the response, and $\bg_i\in\mathcal{G}$ is a $q$-dimensional group or environment index. For example, a two-dimensional ($q=2$) group index might comprise gender and race indicators. If $\bg_i=\bg_{i'}$, then the $i$-th and $i'$-th individuals belong to the same group.

For each group, we consider the following linear model:
\begin{align}
\label{lm}
\E[y_i|\bx_i,\bg_i=g]=\bx_i^\top \bbeta^{(g)},~~\forall g\in\mathcal{G},
\end{align}
where $\bbeta^{(g)}\in\R^p$ denotes the coefficient vector for group $g$. However, data is available only for a subset of groups, denoted as $\mathcal{O}\subseteq \mathcal{G}$. Let $n^{(g)}$ represent the sample size for group $g$. By definition, $n^{(g)}>0$ for each $g\in\mathcal{O}$, and $n^{(g)}=0$ for each $g\notin\mathcal{O}$. Our primary objective is to establish prediction rules for some unseen domains $g\notin\mathcal{O}$.

We propose to organize the coefficient vectors from multiple groups as a high-order tensor and develop new tensor completion methods tailored for domain generalization. Notably, our coefficient tensor presents structured missing patterns, which is different from commonly studied random missing tensor completion scenarios. To this end, we present a novel algorithm named ``TensorDG", which stands for \textbf{Tensor} completion-based algorithm for \textbf{D}omain \textbf{G}eneralization. We further establish the convergence rates of our proposal and show that it is minimax optimal under mild conditions. Additionally, recognizing the diverse requirements of real-world applications, we develop extensions of our core methodology, including the ``TensorTL" approach for transfer learning when the target domain possesses limited samples and a high-dimensional counterpart if the dimension $p$ is larger than $\min_{g\in\mathcal{O}}n^{(g)}$ but $\bbeta^{(g)}$ is sparse. 

We highlight two key features of our proposal. First, it leverages the group structures rather than simply labeling the observed groups as $1,\dots,|\mathcal{O}|$ as in many existing works. This structural information is helpful for understanding the similarity of different domains and further sheds light on devising more explainable and reliable domain generalization methods. Second, our proposal has solid theoretical guarantees and enjoys minimax optimality under mild conditions.  Moreover, by employing the {rank determination techniques \citep{han2022rank}}, practitioners can ascertain the degree to which the model may be misspecified, adding another layer of reliability to the method.%In addition, utilizing the test of low-rank tensors, one can also check whether the model is mis-specified to some extent.

\subsection{Related literature}
\paragraph{Domain generalization.} Existing literature has studied identifying causal features for domain generalization, i.e., the features that are responsible for the outcome but are independent of the unmeasured confounders in each domain.
Identification of causal features has been connected with the estimation of invariant representations \citep{buhlmann2020invariance}. \citet{rojas2018invariant} propose the framework of invariant risk minimization (IRM) in order to find invariant representations across multiple training environments.
However, \citet{kamath2021does} and \citet{choe2020empirical} find that the sample version of IRM can fail to capture the invariance in empirical studies. %Under linear structural causal models, the causal effect identification through the invariant property has been studied in different models \citep{peters2016causal,pfister2019invariant,buhlmann2020invariance}. 
For the purpose of domain generalization, \citet{chen2020domain} propose new estimands with theoretical guarantees under linear structural equation models. \citet{pfister2021stabilizing} investigate so-called stable blankets for domain generalization but the proposed algorithm does not have theoretical guarantees.

Beyond the causal framework, other popular domain generalization methods include Maximin estimator, self-training, and invariant risk minimization. To name a few, distributionally robust optimization \citep{volpi2018generalizing,sagawa2019distributionally} or Maximin estimator \citep{meinshausen2015maximin, guo2023statistical} minimizes the max prediction errors among the training groups. \citet{kumar2020understanding} study the theoretical properties of self-training with gradual shifts. %\citet{pan2010domain} propose to find common latent representations of the source and the target features. 
\citet{baktashmotlagh2013unsupervised} propose a domain invariant projection approach by extracting the information that is invariant across the source and target domains but it lacks theoretical guarantees. \cite{wimalawarne2014multitask,yang2016deep,li2017deeper,feng2021provable} utilize the low-rank matrix or tensor for domain generalization in deep neural networks. However, these methods are either purely empirical or computationally demanding, lacking of statistical optimality guarantee with efficient algorithms. In the realm of invariant predictors, \cite{arjovsky2019invariant} introduce Invariant risk minimization, with extended discussions and elaborations available in \citet{rosenfeld2020risks}, \citet{zhou2022sparse}, and \citet{fan2023environment}. %Additionally, it is noteworthy that several domain adaptation algorithms, such as CORAL \citep{sun2016deep} and DANN \citep{ganin2015unsupervised}, can be {\color{red}seamlessly repurposed} for domain generalization tasks. 
In contrast to our work, the works mentioned above do not consider the structural information of group labels and the generalizability is simply based on model assumptions. Our model leverages the structure of group indices which better explains why and how the model generalizes. It is also possible to verify whether our model assumptions fail or not.

\paragraph{Tensor completion.} 
Our research is also closely related to the tensor estimation and completion, which has significantly advanced in recent years \citep{bi2021tensors}. 
%In the tensor completion literature, most existing works consider the setting where the observed data are directly generated from the target tensor. 
 \citet{montanari2018spectral}  study tensor completion with random missing patterns in the noiseless setting. When having noisy entries, \citet{zhang2019cross} study tensor completion under low-rank assumptions with structural missing. 
 \citet{xia2021statistically} study noisy tensor completion with random missing patterns under low-rank assumptions. The aforementioned two works can be viewed as mean estimation problems among many others, while we aim at estimating the regression coefficients with tensor structures.  
The tensor completion problem can also be rewritten as a tensor regression model whose design consists of indicator functions. \citet{chen2019non} study projected gradient descent for tensor regression and \citet{zhang2020islet} propose a minimax optimal method for low-rank tensor regression with independent Gaussian designs.  \citet{mu2014square} and \citet{raskutti2019convex} study tensor recovery with convex regularizers without and with noises, respectively. From the application perspective, tensor models have been widely used in recommender systems and modeling biomedical image data \citep{adomavicius2010context,zhou2013tensor}.

%Domain generalization can be viewed as an extreme case of transfer learning or few-shot learning as transfer learning has a small batch of samples from the target domain but we have no target samples in the current settings. Transfer learning has been extensively studied in the statistical community recently.

\subsection{Organization and notation}
\label{sec1-4}
In the rest of this paper, we introduce the low-rank tensor model for multi-task regression in Section \ref{sec2}. In Section \ref{sec3}, we present the rationale of the proposed method and introduce the formal algorithm.  We provide theoretical guarantees for the proposed method in Section \ref{sec4}, and discuss extensions of our proposal to transfer learning and to the high-dimensional setting in Section \ref{sec-ext}. In Section \ref{sec-sim}, we demonstrate the numerical performance of our proposal in multiple numerical experiments. In Section \ref{sec-data}, we apply the proposed method to predict the education levels for different ethnic groups. We conclude this paper with discussions in Section~\ref{sec:discussion}.

For a generic matrix $T\in\R^{p_1\times p_2}$, let $\|T\|_2$ denote its spectral norm. Let $\|T\|_{2,\infty}$ denote $\max_{j\leq p_2}\|T_{.,j}\|_2$ and let $\|T\|_{\infty,2}$ denote $\max_{j\leq p_1}\|T_{j,.}\|_2$. For a generic semi-positive definite matrix $\Sig\in\R^{m\times m}$, let $\Lambda_{\max}(\Sig)$ and $\Lambda_{\min}(\Sig)$ denote the largest and smallest singular values of $\Sig$, respectively. Let $\textrm{Tr}(\Sig)$ denote the trace of $\Sig$. For a generic set $A$, let $|A|$ denote the cardinality of $A$. Let $a\vee b$ denote $\max\{a,b\}$ and $a\wedge b$ denote $\min\{a,b\}$. We use $c,c_0,c_1,\dots$ to denote generic constants which can be different in different statements. Let $a_n=O(b_n)$ and $a_n\lesssim b_n$ denote $|a_n/b_n|\leq c$ for some constant $c$ when $n$ is large enough. %Let $a_n=O_P(b_n)$ and $a_n\lesssim_{\P} b_n$ denote $\P(|a_n/b_n|\leq c)\rightarrow 1$ for some constant $c<\infty$. Let $a_n=o_P(b_n)$ denote $\P(|a_n/b_n|>c)\rightarrow 0$ for any constant $c>0$.
%%%%%%%%%%%%%%%%%%%%%
\section{Set-up and Data Generation Model}\label{sec2}
In this section, we outline the basic concepts related to the low-rank tensor model and establish its connection with the domain generalization problem.

\subsection{Low-rank tensor model}
\label{sec2-1}
Invoking Section \ref{sec1-1}, the observed data can be reshaped as $(X^{(g)},\by^{(g)})\in\R^{n^{(g)}\times (p+1)}$,  whose each row corresponds to a sample  $(\bx_i^{\top},y_i)$ with group label $\bg_i=g$ for each $g\in\mathcal{O}$. For each $g\notin\mathcal{O}$, let $\bx_0^{(g)}$ and $y_0^{(g)}$ denote the design and response variables generated from the oracle model for domain $g$. We assume the set of groups is $\mathcal{G}=[p_1]\circ\dots\circ[p_q]$, where ``$\circ$'' denotes the Cartesian product. The categorical group labels can also be encoded as integers without loss of generality.
We assume the following linear models for groups $g\in\mathcal G$: 
\begin{align}
&y_i^{(g)} =( \bx_i^{(g)})^\top \bbeta^{(g)} + \epsilon_{i}^{(g)},~i\in[n^{(g)}],~\forall g\in\mathcal{O},\nonumber\\
&\E[y_0^{(g)}|\bx_0^{(g)}]=(\bx_0^{(g)})^\top \bbeta^{(g)},~\forall g\in\mathcal{G}\setminus\mathcal{O},\label{model2}
\end{align}
where $\epsilon_{i}^{(g)}$ are the independent random noises such that $\E[\eps_i^{(g)}|\bx_i^{(g)}]=0$ for each $g\in\mathcal{O}$. We assume that $\eps_i^{(g)}$ is mutually independent of $\eps_{i'}^{(g')}$ for any $g\neq g'\in\mathcal{O}$, $i\in[n^{(g)}]$, $i'\in[n^{(g')}]$. That is, the noises in different domains are independent. For the unseen groups $g\in\mathcal{G}\setminus\mathcal{O}$, we only make assumptions on the signal part of true models as there are no samples. 

We arrange the regression coefficients $\{\bbeta^{(g)}\}_{g \in \mathcal{G}}$ into a tensor $\bbeta(\mathcal{G}) \in\mathbb{R}^{p\times p_1\times\cdots\times p_q}$ such that
\[
   \{\bbeta(\mathcal{G})\}_{j,i_1,\dots,i_q}=\beta_j^{(i_1,\dots, i_q)}.
\]
In words, the first mode of $\bbeta(\mathcal{G})$ represents the regression coefficients and the remaining $q$ modes represent group indices. We refer to the first mode of $\bbeta(\mathcal{G})$ as the "coefficient mode" and the last $q$ modes as the ``group modes". We write $p_0=p$ for the ease of presentation.

We assume that tensor $\bbeta(\mathcal{G}) $ has Tucker rank $(r_0, r_1, \cdots, r_q)$.
The Tucker rank is defined based on matricization (or matrix unfolding, flattening). Specifically,
the \textit{matricization} $\bM_t[X]$ maps a tensor $X\in\R^{p_1\times \dots\times p_k}$ into a matrix $\bM_t[X] \in \R^{p_t\times(\prod_{1\leq s\neq t\leq k}p_s)}$ such that 
\begin{align}
\label{unfolding}
(\bM_t[X])_{i_t,j}=X_{i_1,i_2,\dots,i_k},~\text{for}~j=1+\sum_{l\leq k,l\neq t}(i_l-1) J_l~\text{and}~J_l=\prod_{1\le m\leq l-1,m\neq t} p_m.
\end{align}
\
The assumption that $\bbeta(\mathcal{G})$ has Tucker rank $(r_0, r_1, \cdots, r_q)$ requires that rank$(\bM_t[\bbeta(\mathcal{G})])= r_t$ for $t=0,\dots,q$, where $r_t$ can be unknown a priori. We illustrate the implications of the low-rankness in the following example.

If $q=2$, the order-3 tensor $\bbeta(\mathcal{G})$ can be unfolded as
\begin{align}
  & \bM_0[\bbeta(\mathcal{G})]=\begin{pmatrix}\bbeta^{(1,1)}&\cdots&\bbeta^{(p_1,p_2)}\end{pmatrix}\in\R^{p\times(p_1p_2)},\nonumber\\
  &\bM_1[\bbeta(\mathcal{G})]=\begin{pmatrix}(\bbeta^{(1,1)})^\top &\dots&(\bbeta^{(1,p_2)})^\top \\
  &\dots&\\
  (\bbeta^{(p_1,1)})^\top &\dots&(\bbeta^{(p_1,p_2)})^\top 
  \end{pmatrix},~~\bM_2[\bbeta(\mathcal{G})]=\begin{pmatrix}(\bbeta^{(1,1)})^\top &\dots&(\bbeta^{(p_1,1)})^\top \\
  &\dots&\\
  (\bbeta^{(1,p_2)})^\top &\dots&(\bbeta^{(p_1,p_2)})^\top \label{unfolding}
  \end{pmatrix}.
\end{align}
The low-rank nature of $\bM_0[\bbeta(\mathcal{G})]$ implies that the coefficient vectors are spanned by a set of $r_0$ basis vectors, allowing each $\bbeta^{(g)}$ to be represented as an $r_0$-dimensional latent score vector in this reduced subspace. Similar low-rank assumptions for linear coefficients have been adopted by \citet{tripuraneni2021provable} in the context of transfer learning. There, the source data is employed to learn the common basis and a limited amount of target data is used to learn the score vectors associated with each domain. However, when dealing with domain generalization, there are no samples available from the target domains and the associated score vector cannot be directly estimated. Fortunately, the low Tucker rank structure suggests that the matrices $\bM_1[\bbeta(\mathcal{G})]$ and $\bM_2[\bbeta(\mathcal{G})]$ also possess a low-rank nature. Such a correlation structure among different groups enlightens the possibility of zero-shot learning. 

While we focus on the Tucker rank, another common metric of tensor rank is the canonical polyadic (CP) rank \citep{hitchcock1927expression}. Let us denote the CP rank of $\bbeta(\mathcal{G})$ as $R(\bbeta(\mathcal{G}))$. It holds that
$\max_{0\leq t\leq q}r_t\leq R(\bbeta(\mathcal{G}))\leq \prod_{t=0}^qr_t/(\max_{0\leq t\leq q}r_t)$. As a result, a tensor with low CP rank will imply a low Tucker rank structure. Hence, we focus on the Tucker rank characterization in this work.

Additionally, we define the mode product as follows. Let $\mathcal{G}'=[h_1]\circ\dots\circ[h_q]$ denote a generic $q$-dimensional index set.
For $E_t\in\R^{h_t\times m_t}$, $t=1,\dots,q$, the $t$-th mode product $\bbeta(\mathcal{G}')\times_t E_t$ is defined as
\[
   \{\bbeta(\mathcal{G}')\times_t E_t\}_{j,i_1,\dots,i_q}=\sum_{s=1}^{h_t} \{\bbeta(\mathcal{G}')\}_{j,i_1,\dots,i_{t-1},s,i_{t+1},\dots,i_q}\{E_t\}_{s,i_t}\in\R^{p\times h_1\times \dots\times h_{t-1}\times m_t\times h_{t+1}\times\dots\times h_q}, 
\]
 for $j\in[p], i_1\in[h_1], ...,  i_q\in[h_q]$. For $E_0\in\R^{p\times m_0}$, the $0$-th mode product is defined as
 \[
    \{\bbeta(\mathcal{G}')\times_0 E_0\}_{j,i_1,\dots,i_q}=\sum_{s=1}^{p} \{\bbeta(\mathcal{G}')\}_{s,i_1,\dots,i_q}\{E_0\}_{s,j}\in\R^{m_0\times h_1\times\dots\times h_q}.
 \]

\subsection{Observed group structures}
\label{sec2-2}
To recover the tensor $\bbeta(\mathcal{G})$, the observed groups $\mathcal{O}$ need to include some crucial elements.  For $t=1,\dots, q$, the \textit{arm} set for mode $t$ is defined as
\begin{align}
\label{def-arm}
 \calA_t=\mathop{\circ}\limits_{1\leq k\neq t\leq q} S_k~\text{such that}~ S_k\subseteq[p_k]~\text{and}~S_1\circ\dots \circ S_{t-1}\circ [p_t]\circ S_{t+1}\circ \dots \circ S_q\subseteq \mathcal{O},
\end{align}
where $\calA_t$ is a $(q-1)$-dimensional set.
In words, for example, if $\bm{a}\in \calA_1$, then $(1,\bm{a}^{\top}),\dots,(p_1,\bm{a}^{\top})$ are all elements of $\mathcal{O}$, i.e., $(1,\bm{a}^{\top}),\dots,(p_1,\bm{a}^{\top})$ are all observed groups. To ease our notation, for $\calA_t$ defined in (\ref{def-arm}), we denote the product $S_1\circ\dots \circ S_{t-1}\circ [p_t]\circ S_{t+1}\circ \dots \circ S_q$ as $\calA_t\circ _t [p_t]$. 
Further, we define the \textit{body} set as
\begin{align}
\label{def-body}
    \Omega=\mathop{\circ}_{1\leq t\leq q} \Omega_t~\text{such that}~ \Omega_t\subseteq[p_t]~~\text{and}~~\Omega\subseteq \mathcal{O}.
\end{align}
In words, the body set $\Omega$ corresponds to a fully observed tensor $\bbeta(\Omega)$ and the set $\Omega_t$ corresponds to the $s$-th coordinate of $\Omega$. 
As a consequence, it holds that
\begin{align}
\label{obs-grps}
        \mathcal{O}\supseteq \Omega\cup(\calA_1\circ_1 [p_1])\cup\dots\cup(\calA_q\circ_q [p_q]).
\end{align}
That is, the observed groups should consist of a body set and $q$ arm sets. {\color{black}Note that $\calA_t$, $\Omega_t$, and $\mathcal{O}$ are all subsets of groups indices which are exclusive of the coordinates of the regression coefficients.}
Notably, our definition of arm set $\calA_t$ is different from the definition in \citet{zhang2019cross}. Specifically, they define the $t$-th arm set, say $\calA'_t$, to be such that $\calA'_t\subseteq \mathop{\circ}\limits_{1\leq s\neq t\leq q}\Omega_s$, which is not needed in our work. As $\calA_t\supseteq \calA'_t$, our proposed method can leverage more observed samples.

\begin{figure}[H]
\centering
\includegraphics[width=0.8\textwidth]{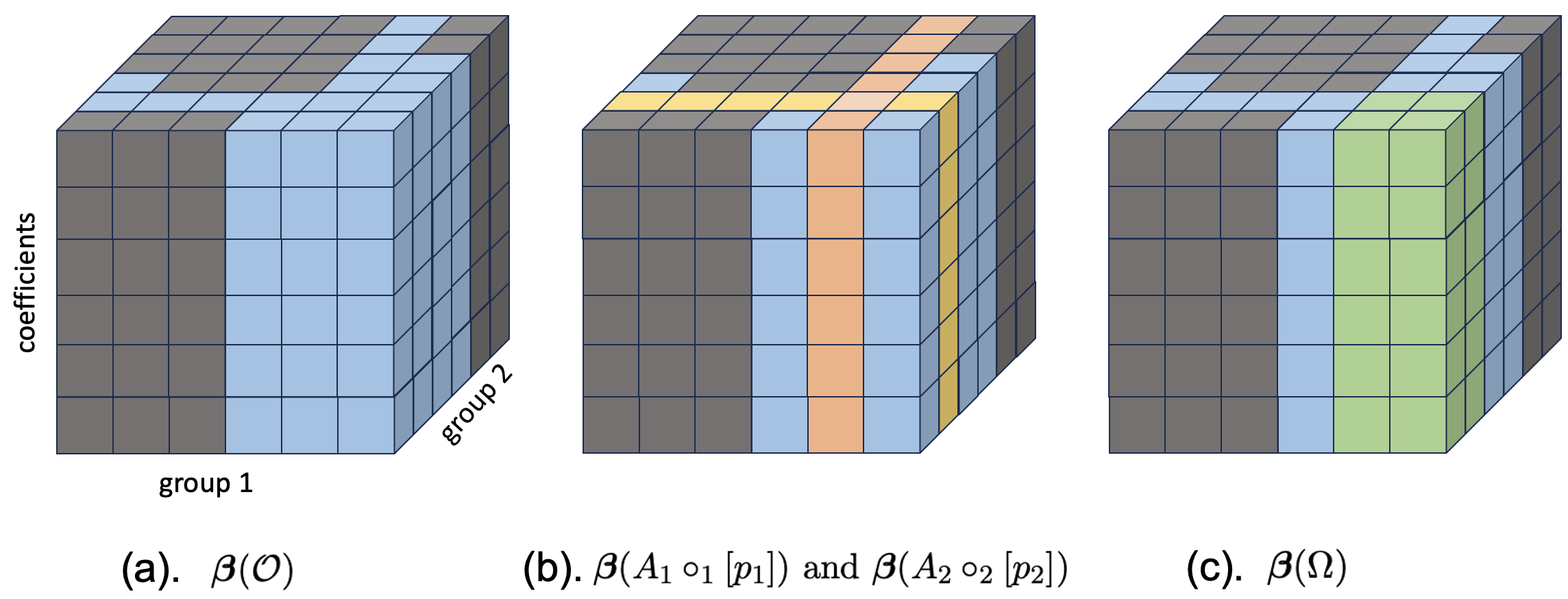}
\caption{The graphical illustration of the coefficient tensor corresponding to the observed set (blue), arm sets (yellow and orange), and body set (green). %{\color{red}The caption need to be updated : $\times_t\rightarrow \circ_t$}
} 
\label{fig:illustration}
\end{figure}

%$$
%\bbeta(A_1 \circ_1 [p_1] ) \text{ and } \bbeta(A_2\circ_2[p_2])
%$$
%$$
%\bbeta(\mathcal{O})
%$$
%As a simple example, suppose that $q=3$, $p_1=p_2=3$, $p_3=2$, and
%\begin{align*}
%    \mathcal{O}=\{(1,1,1),(1,1,2),(1,2,1),(1,3,1),(2,1,1),(3,1,1), (3,3,3)\}.
%\end{align*}
%In this case, we can define
%$\calA_t=(1,1)$ for $t=1,2,3$ and $\Omega=\{1,3\}\circ \{1,3\}\circ \{1\}$.

To summarize, the missing data pattern described in (\ref{obs-grps}) is structural, distinguishing it from missing at random. 
\citet{xia2021statistically} studies minimax optimal tensor completion methods with missing completely at random (MCAR) among many others. The MCAR assumption cannot be directly applied in our problem, given that $\bbeta(\mathcal{G})$ has no missingness in the $0$-th mode. %Another significant feature of $\bbeta(\mathcal{G})$ is that $p$ can be much larger than $\max_{1\leq t\leq q}p_t$ in many applications and $r_0$ can also be larger than $\max_{1\leq t\leq q}r_t$. {\color{red}Hence, using the max rank $\max_{0\leq t\leq q}r_q$ to approximate each individual rank may not lead to sharp analysis, making our results different from those in \citet{montanari2018spectral} and  \citet{barak2016noisy}.}

\section{Method}
\label{sec3}
In this section, we introduce the proposed method for domain generalization.
We first outline the rationale of our proposal based on Tucker decomposition in Section \ref{sec3-1}. The formal algorithm is provided in Section \ref{sec3-2}.
\subsection{Rationale from Tucker decomposition}
\label{sec3-1}

Recall the definition of mode product at the end of Section \ref{sec2-1}.
If $\bbeta(\Omega)$ has Tucker rank $(r_0, r_1\dots, r_t)$, then the Tucker decomposition of $\bbeta(\mathcal{G})$ can be written as
\begin{align}
\label{eq-decomp1}
    \bbeta(\mathcal{G})=\bbeta(\Omega)\times_0R_0\times_1 R_1\times_2 \dots\times_q R_q\in\R^{p\times p_1\times\dots\times p_q},
\end{align}
where $R_0\in\R^{p\times p}$ and $R_t\in\R^{\omega_t\times p_t}$, $t=1,\dots,q$ are computed based on $\bbeta(\calA_t\circ_t[p_t])$ as in the forthcoming (\ref{def-Rt}). Equation (\ref{eq-decomp1}) implies that if $\bbeta(\Omega)$ is a sufficient dimension reduction of $\bbeta(\mathcal{G})$ and $R_t$ are well-defined, then the full tensor can be recovered by using a subset of groups $\mathcal{O}$ satisfying (\ref{obs-grps}), which only involves a small number of groups as illustrated in Figure \ref{fig:illustration} in contrast to a total of $\prod_{t=1}^qp_t$ groups.

We now provide more details about $R_t$ in (\ref{eq-decomp1}).
Indeed, $R_t$ can be identified based on the joint measurements $B_t^{(jo)} =\bM_t^\top[\bbeta(\calA_t\circ_t \Omega_t)]\in\R^{(|\calA_t|p)\times \omega_t}$ and the arm measurements $B_t^{(ar)}=\bM_t^\top [\bbeta( \calA_t\circ_t [p_t])]\in\R^{(|\calA_t|p)\times p_t}$. For $t=0$, especially, we define $B_0^{(jo)}=B_0^{(ar)}=\bM_0^{\top}[\bbeta(\mathcal{O})]\in\R^{|\mathcal{O}|\times p}$. If rank$(B_t^{(jo)})=r_t$ for $t=0,\dots,q$, then (\ref{eq-decomp1}) holds with 
\begin{align}
\label{def-Rt}
  R_t=(B_t^{(jo)})^{\dagger}B_t^{(ar)}\in\R^{ \omega_t\times p_t},
\end{align}
where the notation $A^{\dagger}$ denotes the pseudo-inverse of a matrix $A$. In fact, $R_t$ also lives in a low-dimensional subspace.
Consider the SVD of $B_t^{(jo)}=U_t\Lambda_tV_t^\top $.
By (\ref{def-Rt}), we have the following relationships
\begin{align}
\label{def-Rt2}
   V_tV_t^\top R_t=V_t\Lambda_t^{-1}U_t^\top B_t^{(ar)}=R_t, ~t=0,\dots,q.
\end{align}
That is, $R_t$ lies on the linear subspace spanned by $V_t$. Together with (\ref{eq-decomp1}), we have
\begin{align}
\label{eq-decomp2}
\bbeta(\mathcal{G})=\bbeta(\Omega)\times_0 V_0V_0^\top \times_{t=1}^qV_tV_t^\top R_t=(\bbeta(\Omega)\times_{t=0}^qV_t)\times_{t=0}^q \Gam_t,
\end{align}
where $\Gam_0=V_0^\top \in\R^{r_0\times p_0}$ and $\Gam_t=V_t^\top R_t\in\R^{r_t\times p_t}$, $t=1,\dots,q$.
Comparing (\ref{eq-decomp2}) with (\ref{eq-decomp1}), we see that to recover the whole tensor, it suffices to estimate $\bbeta(\Omega)\times_{t=0}^q V_t\in\R^{r_0\times r_1\dots\times r_q}$ and $\Gam_t\in\R^{r_t\times p_t}$ for $t=0,\dots,q$, which has degree of freedom  $\prod_{t=0}^qr_t+\sum_{t=0}^q(p_t-r_t)r_t$. Indeed, $\bbeta(\Omega)\times_{t=0}^q V_t$ is always referred to as the core tensor \citep{zhang2018tensor} as it is a smallest possible tensor which spans the subspaces of each mode and $\Gam_t$ can be viewed as multiplying coefficients for the $t$-th direction. We will propose an algorithm to estimate the core tensor and $\{\Gam_t\}_{t=0}^q$ for tensor completion in the next subsection.
\begin{remark}
Model (\ref{eq-decomp1}) is also related to the tensor factor models \citep{han2020tensor,chen2022factor}, which have been studied in high-dimensional tensor time series. Using the terminology in factor models, $\Gam_t$ are the loading matrices and the core tensor $\bbeta(\Omega)\times_{t=0}^qV_t$ is the tensor of factors. In the time series applications, the observed data forms a complete tensor which is different from our setting. 
\end{remark}

%\begin{remark}
%Model (\ref{eq-decomp1}) is also related to the tensor factor models \citep{han2020tensor,chen2022factor}, which have been studied for high-dimensional tensor time series. Using the terminology in factor models, $\Gam_t$ are the loading matrices and the core tensor $\bbeta(\Omega)\times_{t=0}^qV_0$ is the tensor of factors. In the time series applications, the observed data comprise a complete tensor which is different from our setting. 
%\end{remark}

\subsection{Proposed algorithm}
\label{sec3-2}
We now devise an algorithm to estimate $\bbeta(\mathcal{G})$ based on (\ref{eq-decomp2}). Our proposal has three main steps. The first step is to estimate the low-dimensional subspace $V_t$ for $t=0,\dots,q$. Then we estimate $\Gam_t$ based on (\ref{def-Rt}) and (\ref{def-Rt2}). Finally, we assemble the estimated tensor based on (\ref{eq-decomp2}). The proposal, named as TensorDG, is presented in Algorithm \ref{alg-dg}.

       \begin{algorithm}[H]
\textbf{Input}: $\{X^{(g)},\by^{(g)}\}_{g\in \mathcal{O}}$, body set $\Omega$, arm sets $\calA_t$, $t=1,\dots,q$.

\textbf{Output}: $\widehat{\bbeta}(\mathcal{G})\in\R^{p\times p_1\times\dots\times p_q}$.

\textbf{Step 0: Sample splitting}. For each $g\in\mathcal{O}$, split the sample into two disjoint folds such that $I_1^{(g)}\cup I_2^{(g)}=[n^{(g)}]$ and $|I_1^{(g)}|\approx |I_2^{(g)}|$. Let $(\widetilde{X}^{(g)},\tilde{\by}^{(g)})\in\R^{ |I_1^{(g)}|\times(p+1)}$ and $(\mathring{X}^{(g)},\mathring{\by}^{(g)})\in\R^{ |I_2^{(g)}|\times(p+1)}$ denote the observations within first and second folds, respectively.

\textbf{Step 1: Estimation of the rank and basis for each mode}. For $t=0,\dots,q$, estimate $\tilde{r}_t$ and $\widetilde{V}_t$ via Algorithm \ref{alg-svd} with input $\{\widetilde{X}^{(g)},\tilde{\by}^{(g)}\}_{g\in\mathcal{O}}$.

\textbf{Step 2: Estimate $\Gam_t=V_t^\top R_t$}. 
For each $g\in\mathcal{O}$, compute $\tilde{\bbeta}^{(g)}$ based on data in $I_1^{(g)}$, $g\in\mathcal{O}$ via (\ref{beta-tilde}). Compute
the OLS estimate $\mathring{\bbeta}^{(g)}=\{(\mathring{X}^{(g)})^\top\mathring{X}^{(g)} \}^{-1}(\mathring{X}^{(g)})^{\top}\mathring{\by}^{(g)}$.

 \For{mode $t=1,\dots, q$}{
Unfold $\widetilde{B}^{(jo)}_t=\bM_t^\top [\widetilde{\bbeta}(\calA_t\circ_t \Omega_t)]\in
R^{a_tp\times \omega_t}$, $\mathring{B}^{(jo)}_t=\bM_t^\top [\mathring{\bbeta}(\calA_t\circ_t \Omega_t)]\in\R^{a_tp\times\omega_t}$, and $\mathring{B}^{(ar)}_t=\bM_t^{\top}[\mathring{\bbeta}(\calA_t\circ_t [p_t])]\in\R^{a_tp\times p_t}$. Compute
\begin{align}
\label{eq-Rhat-ss}
\widehat{\Gam}_t=(\widetilde{V}_t^\top (\widetilde{B}^{(jo)}_t)^\top \mathring{B}^{(jo)}_t\widetilde{V}_t)^{-1}(\widetilde{B}_t^{(jo)}\widetilde{V}_t)^\top \mathring{B}_{t}^{(ar)}\in\R^{\tilde{r}_t\times p_t}.
\end{align}
}
For $t=0$, unfold $\widetilde{B}_0=\bM_0^{\top}[\tilde{\bbeta}(\mathcal{O})]\in\R^{|\mathcal{O}|\times p}$, $\mathring{B}_0=\bM_0^{\top}[\mathring{\bbeta}(\mathcal{O})]\in\R^{|\mathcal{O}|\times p}$ and compute $\widehat{\Gam}_0=(\widetilde{V}_0^\top \widetilde{B}_0^\top \mathring{B}_0\widetilde{V}_0)^{-1}(\widetilde{B}_0\widetilde{V}_0)^\top \mathring{B}_{0}\in\R^{\tilde{r}_0\times p_0}.$

\textbf{Step 3: Tensor completion}.  Compute
        \begin{align*}
        \widehat{\bbeta}(\mathcal{G})&=(\mathring{\bbeta}(\Omega)\times_{t=0}^q\widetilde{V}_t)\times_{t=0}^q\widehat{\Gam}_t.
        \end{align*}
  \caption{TensorDG: Domain Generalization via Tensor Completion}
  \label{alg-dg}
\end{algorithm}
Algorithm \ref{alg-dg} starts with a sample-splitting step, which is mainly for technical convenience. Similar approaches have also been used in existing works to derive theoretical guarantees for tensor estimation and completion \citep{zhang2020islet}. %We evaluate the effects of sample splitting in Section \ref{sec-sim}.

In Step 1, Algorithm \ref{alg-dg} estimates $V_t$ by $\widetilde{V}_t$ using half of the samples via Algorithm \ref{alg-svd}. Algorithm \ref{alg-svd} is motivated by the fact that  $V_t$ is the column space of $\Theta_t=\{\bM_t[\bbeta(\mathcal{C}_t\circ_t \Omega_t)]\}^\top \bM_t[\bbeta(\mathcal{C}_t\circ_t \Omega_t)]/|\mathcal{C}_t|$ for $\mathcal{C}_t=\Omega_{-t}\cup \calA_t$. Hence, we find $V_t$ as the column space of $\Theta_t$, $t=1,\dots,q$.
In Algorithm \ref{alg-svd}, the proposed estimate $\widetilde{\Theta}_t$ of $\Theta_t$  is based on the least square estimates of $\bbeta(\mathcal{C}_t\circ_t \Omega_t)$ and the last term of its expression further corrects the bias caused by the product of OLS estimates. The rank $\tilde{r}_t$ is estimated as the number of significantly nonzero eigenvalues of $\widetilde{\Theta}_t$. The threshold level $\lambda_t$ is chosen based on the concentration properties of $\widetilde{\Theta}_t$. Determination of the tensor rank can also be based on information criteria \citep{han2022rank}.

In Step 2, we obtain estimates of $\bbeta(\Omega)$ based on two disjoint sets of samples. Then we estimate $\Gam_t$ by $\widehat{\Gam}_t$. For $t=0$, a simple estimate of $\Gam_0=V_0^\top $ is $\widetilde{V}_0^\top $. However, the convergence rate of $\widetilde{V}_0$ depends on the eigen-gap condition critically (Lemma \ref{thm1}). Hence, we propose $\widehat{\Gam}_0$ whose convergence rate requires weaker regularity conditions. In Step 3, the whole tensor is assembled according to (\ref{eq-decomp2}) with the aforementioned estimates.  

\begin{algorithm}[H]
\textbf{Input}:  $\{\widetilde{X}^{(g)},\tilde{\by}^{(g)}\}_{g\in\mathcal{O}}$. 

\textbf{Output}: Estimated rank $\tilde{r}_t$ and basis $\widetilde{V}_t\in\R^{\omega_t\times\tilde{r}_t}$.

For each $g\in \mathcal{O}$, let $\tilde{n}^{(g)}=|I_1^{(g)}|$ and compute $\widetilde{\Sig}^{(g)}=(\widetilde{X}^{(g)})^\top\widetilde{X}^{(g)}/\tilde{n}^{(g)}$,
\begin{align}
\tilde{\bbeta}^{(g)}&=\{\widetilde{\Sig}^{(g)}\}^{-1}(\widetilde{X}^{(g)})^\top \tilde{\by}^{(g)}/\tilde{n}^{(g)}~~\text{and}~~\tilde{\sig}^{2,(g)}=\frac{\|\tilde{\by}^{(g)}-\tilde{X}^{(g)}\tilde{\bbeta}^{(g)}\|_2^2}{\tilde{n}^{(g)}-p}.\label{beta-tilde}
\end{align}

- If $t=0$, let $\widetilde{B}_0=\bM_0[\tilde{\bbeta}(\mathcal{O})]\in\R^{|\mathcal{O}|\times p}$.
Denote the following eigenvalue decomposition 
\[
\widetilde{\Theta}_{0}=\frac{1}{|\mathcal{O}|}\widetilde{B}_{0}^\top \widetilde{B}_{0}-\frac{1}{|\mathcal{O}|}\sum_{g\in\mathcal{O}}(\widetilde{\Sig}^{(g)})^{-1}\frac{\tilde{\sig}^{2,(g)}}{\tilde{n}^{(g)}}=\mathring{V}_{0}\mathring{\Lambda}_{0}\mathring{V}_{0}^\top \in\R^{p\times p}.
\]
Let $\tilde{r}_0=\sum_{k=1}^{p} \mathbbm{1}((\mathring{\Lambda}_{0})_{k,k}\geq \lambda_{0})$ and $\widetilde{V}_{0}=\{\mathring{V}_{0}\}_{.,1:\tilde{r}_0}\in\R^{p\times \tilde{r}_0}$, where  $\lambda_0=C\sqrt{\|\widetilde{\Theta}_0\|_2(p+\log \bar{n})/(\bar{n}|\mathcal{C}_t|)}$ for $\bar{n}=\sum_{g\in\mathcal{O}}\tilde{n}^{(g)}/|\mathcal{O}|$.

- If $t\geq 1$, let $\widetilde{B}_t=\bM_t[\tilde{\bbeta}(\mathcal{C}_t\circ_t \Omega_t)]\in\R^{|\mathcal{C}_t|p\times \omega_t}$. Define
\[
\widetilde{\Theta}_t=\frac{1}{|\mathcal{C}_t|}\widetilde{B}_t^\top \widetilde{B}_t-\frac{1}{|\mathcal{C}_t|}\textup{Diag}(\tilde{\bm{v}})=\mathring{V}_t\mathring{\Lambda}_t\mathring{V}_t^\top \in\R^{\omega_t\times\omega_t},
\]
where $\textup{Diag}(\tilde{\bm{v}})$ is a diagonal matrix with $\{\textup{Diag}(\tilde{\bm{v}})\}_{j,j}=\tilde{v}_j$ and $\tilde{\bm{v}}_j=\sum_{g_{-t}\in \mathcal{C}_t, g_t=(\Omega_t)_j}\textup{Tr}(\{\widetilde{\Sig}^{(g)}\}^{-1})\frac{\tilde{\sig}^{2,(g)}}{\tilde{n}^{(g)}}$.
Let $\tilde{r}_t=\sum_{k=1}^{\omega_t} \mathbbm{1}((\mathring{\Lambda}_t)_{k,k}\geq \lambda_t)$ and $\widetilde{V}_t=\{\mathring{V}_t\}_{.,1:\tilde{r}_t}\in\R^{\omega_t\times \tilde{r}_t}$, where $\lambda_t=C\sqrt{\|\widetilde{\Theta}_t\|_2(\omega_t+\log \bar{n})/(\bar{n}|\mathcal{C}_t|)}$.

\caption{SVD for mode $t$.}
\label{alg-svd}
\end{algorithm}

%There are two main differences between our method and the existing tensor completion methods, say, the Cross method in \citet{zhang2019cross}, which is also based on decomposition (\ref{eq-decomp1}). 
Our method exhibits two primary distinctions from existing tensor completion methods, such as the Cross method described by \citet{zhang2019cross}, which also employs decomposition (\ref{eq-decomp1}).
First, in the standard tensor completion problem, {\color{black}each element of the oracle tensor is observed directly, with each observation being independent and unbiased. In contrast, in the domain generalization problem under consideration, we need to fit regression models for each domain. The produced OLS estimates therefore exhibit correlated errors, which need to be calibrated as detailed in Algorithm \ref{alg-svd}.}  Second, for our tensor $\bbeta(\mathcal{G})$, the 0-th mode is not exchangeable with other $q$ (group) modes. This distinction arises because $\bbeta^{(g)}$  represents the smallest unit of interest in domain generalization, and it is either observed or missed in its entirety. Consequently, the dimension reduction approach we adopt for the 0-th mode differs from the strategies employed for the remaining $q$ modes.
%This is because $\beta^{(g)}$ is the smallest unit we are interested in for domain generalization and $\beta^{(g)}$ is observed or unobserved as a whole. Hence, we perform dimension reduction for the 0-th mode differently from that for other $q$ modes. Theoretically, we derive new error bounds to demonstrate TensorDG's performance in domain generalization.

\section{Theoretical Properties}
\label{sec4}
In this section, we provide theoretical guarantees for Algorithm \ref{alg-dg}. We first state the main assumptions.

\begin{condition}[Overall structure]
\label{cond1}
The tensor $\bbeta(\mathcal{G})$ has Tucker rank $(r_0, r_1,\dots, r_q)$ and $q$ is fixed. 
Moreover, $\bM_t[\bbeta(\Omega)]$ and $\bM_t[\bbeta(\calA_t\circ_t \Omega_t)]$ both have rank $r_t$ for $t=1,\dots,q$ and $\bM_0[\bbeta(\Omega)]$ has rank $r_0$.
\end{condition}
 Condition \ref{cond1} assumes that $\bbeta(\Omega)$ is a sufficient dimension reduction of $\bbeta(\mathcal{G})$. The condition that $\bM_t[\bbeta(\calA_t\circ_t \Omega_t)]$ has rank $r_t$ guarantees that $R_t$ in (\ref{def-Rt}) is well-defined. As Condition \ref{cond1} can be violated in practice, we will discuss the model diagnostics at the end of Section \ref{sec4}.

\begin{condition}[Distribution of observed data]
\label{cond2}
For each $g\in\mathcal{O}$, $\bx_i^{(g)}$, $i=1,\dots,n^{(g)}$, is independent sub-Gaussian with mean zero and covariance matrix $\Sig^{(g)}$, where $c_1\leq \min_{g\in\mathcal{G}}\Lambda_{\min}(\Sig^{(g)})\leq \max_{g\in\mathcal{G}}\Lambda_{\max}(\Sig^{(g)})\leq c_2$ for some positive constants $c_1$ and $c_2$.  For each $g\in\mathcal{O}$, the noise $\epsilon_i^{(g)}$ is independent sub-Gaussian with mean zero and variance $\sigma^{2,(g)}$. Moreover, $n^{(g)}\asymp n$ for all $g\in\mathcal{O}$.
\end{condition}

 Condition \ref{cond2} assumes sub-Gaussian designs and sub-Gaussian errors. It is worth highlighting that heterogeneous distributions for both $\bx_i^{(g)}$ and $\eps_i^{(g)}$ are allowed. The assumption that $n^{(g)}\asymp n$ is a simplified scenario for technical convenience and has been commonly considered in the multi-task learning literature \citep{guo2011joint, tripuraneni2021provable}.

We consider the scenario where $(p_t,r_t)$, $t=0,\dots,q$ can all go to infinity but $p\leq c_1n$ for some small enough constant $c_1$. This low-dimensional assumption guarantees the regularity of the least square estimates of each group. In Section \ref{sec-ext}, we discuss possible extensions of the proposed methods to the high-dimensional setting.
For $t=1,\dots,q$, let $\Theta_t=\{\bM_t[\bbeta(\mathcal{C}_t\circ_t \Omega_t)]\}^\top \bM_t[\bbeta(\mathcal{C}_t\circ_t \Omega_t)]/|\mathcal{C}_t|\in\R^{\omega_t\times\omega_t}$ and $\Theta_0=\{\bM_0[\bbeta(\mathcal{O})]\}^\top \bM_0[\bbeta(\mathcal{O})]/|\mathcal{O}|\in\R^{p\times p}$. For $t=0,\dots,q$, define
\begin{align}
  &e(\Theta_t)=\min_{0\leq r\leq r_t}\{\lambda_{r}(\Theta_t)-\lambda_{r+1}(\Theta_t)\}\label{Theta},
\end{align}
where $e(\Theta_t)$ denotes the smallest eigen-gap for matrix $\Theta_t$ with the convention that $\lambda_0(\Theta_t)=\infty$.

\begin{condition}[Eigenvalue conditions]
\label{cond-eigen}
For $\Theta_t$ defined in (\ref{Theta}), assume that $e_0\leq \min_{0\leq t\leq q}e(\Theta_t)$, $e_*\leq \min_{0\leq t\leq q}\Lambda_{r_t}(\Theta_t)\leq \max_{0\leq t\leq q}\Lambda_{\max}(\Theta_t)\leq e^*$,  and $e_*\leq \min_{1\leq t\leq q}\Lambda_{r_t}((B^{(jo)}_t)^\top B_t^{(jo)}/|\calA_t|)\leq \max_{1\leq t\leq q}\Lambda_{\max}((B^{(jo)}_t)^\top B_t^{(jo)}/|\calA_t|)\leq e^*$ for some positive constants $e_*$ and $e^*$.
\end{condition}
Condition \ref{cond-eigen} requires the so-called eigen-gap condition of $\Theta_t$. This condition is needed for estimating $V_t$ as an application of sin$\Theta$ theorem \citep{yu2015useful}. 

\subsection{Upper bounds for the generalization errors}
\label{sec4-1}
We first establish the estimation accuracy of subspace estimation in Step 1 of Algorithm \ref{alg-dg}. Let $\omega_t=|\Omega_t|$ and $a_t=|\calA_t|$ for $t=1,\dots,q$.
\begin{lemma}[Subspace estimation for each mode]
\label{thm1}
Suppose that Conditions \ref{cond1}, \ref{cond2}, and \ref{cond-eigen} hold. If $p(\log|\mathcal{O}|+\log n)\leq c_1n$ and $\omega_t\leq c_1|\mathcal{C}_t|n$ with small enough constant $c_1$, then for any $1\leq t\leq q$,
\begin{align*}
 &  \|\widetilde{V}_t-V_t\|_2\leq \frac{c_2}{e_0}\sqrt{\frac{e^*(\omega_t+\log n)}{n|\mathcal{C}_t|}},~~\tilde{r}_t=r_t
   \end{align*}
 with probability at least $1-\exp\{-c_3\log n\}$. Moreover, with probability at least $1-\exp\{-c_3\log n\}$, 
\[
  \|\widetilde{V}_0-V_0\|_2\leq \frac{c_2}{e_0}\sqrt{\frac{e^*(p+\log n)}{n|\mathcal{O}|}},~~\tilde{r}_0=r_0.
\]
 \end{lemma}

Lemma \ref{thm1} provides the convergence rate of $\widetilde{V}_t$ for $V_t$, $t=0,\dots,q$ under given conditions. The quantity $e_0$ in the denominator shows the effect of eigen gap.  For each $1\leq t\leq q$, we use $n|\mathcal{C}_t|$ samples to estimate $V_t$. As $|\mathcal{C}_t|\geq \prod_{s\neq t}\omega_{s}$, the condition $\omega_t\lesssim n|\mathcal{C}_t|$ is mild.  The condition $p(\log|\mathcal{O}|+\log n)\leq c_1n$ guarantees that  $\min_{g\in\mathcal{O}}\Lambda_{\min}(\widetilde{\Sig}^{(g)})$ is bounded away from zero with high probability. %For the estimation of $\widetilde{V}_0$, the result is analogous to the results for $1\leq t\leq q$ by noticing that $C_0=\mathcal{O}$. 

%%%%%%%%%%%%%
In the following lemma, we decompose the domain generalization errors of $\widehat{\bbeta}(\mathcal{G})$. For a generic tensor $\bm{B}$, let $\|\bm{B}\|_{\ell_2}$ denote its vectorized $\ell_2$-norm. For $t=0,\dots,q$, let $\widetilde{\Gam}_t=(V_t^\top \widetilde{V}_t)^{-1}\Gam_t$. Let $C_R=\max_{1\leq t\leq q}\|R_t\|_{2,\infty}\vee 1$ and $\bar{C}_R=\max_{1\leq t\leq q}\|R_t\|_2$. Let $\bar{\lambda}_{\mathcal{G}}=\|\bM_0[\bbeta(\mathcal{G})]\|_2$, $\bar{\lambda}_{\mathcal{G}_{-t}\circ_t\Omega_t}=\|\bM_t[\bbeta(\mathcal{G}_{-t}\circ_t\Omega_t)]\|_2$, and $\bar{\lambda}_{g_{-t}\circ_t\Omega_t}=\|\bM_t[\bbeta(g_{-t}\circ_t\Omega_t)]\|_2$.
\begin{lemma}[Decomposition of generalization errors]
\label{lem-decomp}
Assume Conditions \ref{cond1}, \ref{cond2}, and \ref{cond-eigen}.  Assume that $p(\log|\mathcal{O}|+\log n)\leq c_1n$, $\omega_t\leq c_1 |\mathcal{C}_t|n$, and $r_t\leq c_1na_t$, $t=1,\dots,q$  for some small enough constant $c_1$. Then we have

\textup{(Frobenius norm)}
{\small
\begin{align}
\|\widehat{\bbeta}(\mathcal{G})-\bbeta(\mathcal{G})\|_{\ell_2}&\lesssim  \sum_{t=1}^q\bar{\lambda}_{\mathcal{G}_{-t}\circ_t\Omega_t}\sqrt{r_t}\|\widehat{\Gam}_t-\widetilde{\Gam}_t\|_2+\bar{\lambda}_{\mathcal{G}}\sqrt{r_0}\|\widehat{\Gam}_0-\widetilde{\Gam}_0\|_2+\bar{C}_R^{q}\sqrt{\frac{\prod_{t=0}^{q}r_t+\eta}{n}}+rem_{\textup{frob}}, \label{re-frob}
\end{align}
}
for $rem_{\textup{HS}}$ defined in (A.10) with probability at least $1-\exp\{-c_2\log n\}-c_3\exp\{-c_4\eta\}$.

\textup{(Max norm)}
\begin{align}
\max_{g\in\mathcal{G}}\|\hat{\bbeta}^{(g)}-\bbeta^{(g)}\|_2&\lesssim  \max_{g\in\mathcal{G}}\sum_{t=1}^q\bar{\lambda}_{g_{-t}\circ_t\Omega_t}\|\widehat{\Gam}_t-\widetilde{\Gam}_{t}\|_{2,\infty}+\max_{g\in\mathcal{G}}\|\bbeta^{(g)}\|_2 \|\widehat{\Gam}_0-\widetilde{\Gam}_0\|_2\nonumber\\
&\quad+C_R^q\sqrt{\frac{r_0+\log|\mathcal{G}|+\eta}{n}}+rem_{\max},\label{re-max}
\end{align}
for $rem_{\max}$ defined in (A.15) with probability at least $1-\exp\{-c_2\log n\}-c_3\exp\{-c_4\eta\}$. 
\end{lemma}
In Lemma \ref{lem-decomp}, we consider two error bounds for the estimated tensor $\widehat{\bbeta}(\mathcal{G})$. The first one (\ref{re-frob}) is in element-wise $\ell_2$-norm, which gives an overall characterization of all the groups. This norm has been widely considered in the existing literature but it does not distinguish the $q$ group modes from the coefficient mode. For the purpose of domain generalization, we also consider the maximum of group-wise $\ell_2$-norm (\ref{re-max}), which demonstrates the generalization accuracy for each group.
In both results, we decompose the error of generalization into three main sources. The first component is the estimation error of $\widehat{\Gam}_t$ for $t=1,\dots,q$. The second component comes from the estimation of $\widehat{\Gam}_0$, which is the dimension reduction for the $0$-th mode and has no missingness. The third term comes from the noise in the estimated core tensor. The remainder terms $rem_{\textup{frob}}$ and $rem_{\max}$ are high-order terms given in the supplements which are not dominant under mild conditions. Based on these decompositions, we provide formal upper bounds for our proposal. %

\begin{theorem}[Domain generalization bounds in element-wise $\ell_2$-norm]
\label{thm-dg}
Assume Conditions \ref{cond1}, \ref{cond2}, and \ref{cond-eigen}. Suppose that $p(\log|\mathcal{O}|+\log n)\leq c_1n$, $\omega_t\leq c_1 |\mathcal{C}_t|n$, and $r_t\leq c_1na_t$, $t=1,\dots,q$ for some small enough constant $c_1$. In addition, if $\bar{C}_R^{q-2}\sum_{t=1}^q\sqrt{\frac{\bar{C}_R^2r_t+p_t+\bar{C}_R^2\log n}{e_*na_t}}=O(1)$, then for any fixed $\eta\lesssim \log n$, with probability at least $1-\exp\{-c_2\log n\}-c_3\exp\{-c_4\eta\}$ 
\begin{align}
&\sqrt{\sum_{g\in\mathcal{G}}\frac{1}{n^{(g)}}\|X^{(g)}(\hat{\bbeta}^{(g)}-\bbeta^{(g)})\|_2^2}\vee \|\hat{\bbeta}(\mathcal{G})-\bbeta(\mathcal{G})\|_{\ell_2} \lesssim \sum_{t=1}^q\lambda_{\mathcal{G}_{-t}\circ_t\Omega_t}\sqrt{\frac{(\bar{C}_R^2r_t+p_t+\bar{C}_R^2\eta)r_t}{a_tne_*}}\nonumber\\
&\quad\quad\quad+\bar{\lambda}_{\mathcal{G}}\sqrt{\frac{(p+\eta)r_0}{|\mathcal{O}|ne_*}}+\bar{C}_R^{q}\sqrt{\frac{\prod_{t=0}^{q}r_t+\eta}{n}}.\label{eq1-frob}
\end{align}

%\textup{(Max norm)}
%(ii) If $\max_{1\leq t\leq q}(C_R^2r_t+\log p_t)/a_t\leq e_*n^{\alpha}$ and  $\frac{q\bar{\lambda}_{\Omega}C_R^{q-2}}{\max_{g\in\mathcal{G}}\min_t \bar{\lambda}_{g_{-t}\circ \Omega_t}}+\frac{q\bar{\lambda}_{\Omega}C_R^{q-1}}{\max_{g\in\mathcal{G}}\|\bbeta^{(g)}\|_2}\leq n^{(1-\alpha)/2}$ for some $\alpha\leq 1$, then with probability at least $1-\exp\{-c_1\min_{0\leq t\leq q}(r_t\wedge \log p_t)\}$,
%\begin{align}
% \max_{g\in\mathcal{G}}\|\hat{\bbeta}^{(g)}-\bbeta^{(g)}\|_2&\lesssim  \max_{g\in\mathcal{G}}\sum_{t=1}^q\bar{\lambda}_{g_{-t}\circ \Omega_t}\sqrt{\frac{C_R^2r_t+\log p_t}{a_tne_*}}+\max_{g\in\mathcal{G}}\|\bbeta^{(g)}\|_2\sqrt{\frac{pr_0}{|\mathcal{O}|ne_*}}+C_R^q\sqrt{\frac{r_0+\log|\mathcal{G}|}{n}}.\label{eq1-max}
%\end{align}
\end{theorem}
Theorem \ref{thm-dg} provides an upper bound for the estimation errors in all the environments including the unseen ones.  To further understand this result, if $\bar{C}_R\leq c<\infty$ and $\eta\lesssim\min_{0\leq t\leq q}p_t\wedge (\prod_{t=0}^q r_t)$, then the upper bound in (\ref{eq1-frob}) can be rewritten as
\begin{align}
\label{eq2-frob}
&\sqrt{\frac{\prod_{t=0}^q r_t}{n}}+\bar{\lambda}_{\mathcal{G}}\sqrt{\frac{pr_0}{|\mathcal{O}|ne_*}}+\sum_{t=1}^q\bar{\lambda}_{\mathcal{G}_{-t}\circ_t\Omega_t}\sqrt{\frac{r_tp_t}{na_te_*}}.
\end{align}
It is known that the degree of freedom for a tensor with rank $(r_0,r_1,\dots,r_q)$ is $\prod_{t=0}^qr_t+\sum_{t=0}^qr_tp_t$. Loosely speaking, the upper bound in (\ref{eq2-frob}) shows that the estimation error $\widehat{\bbeta}(\mathcal{G})$ is equivalent to estimating $\prod_{t=0}^qr_t+\sum_{t=0}^qr_tp_t$ parameters with the observed samples.
From the perspective of tensor completion, Theorem 2 in \citet{zhang2019cross} considers the case that each observed element in the tensor has one independent sample and its upper bound involves the magnitude of empirical noises.
The result (\ref{eq2-frob}) is more general in the sense that in the current case, each observed element in the tensor corresponds to a regression model with $n^{(g)}\asymp n$ independent samples.

The condition $\bar{C}_R^{q-2}\sum_{t=1}^q\sqrt{\frac{\bar{C}_R^2r_t+p_t+\bar{C}_R^2\log n}{e_*na_t}}=O(1)$ guarantees that the high-order term $rem_{\textup{frob}}$ is dominated by other terms. The conditions essentially require that $q$ and $\bar{C}_R$ should grow to infinity relatively slow.
%It is not hard to see from the proof that these conditions can be further weakened if we perform another sample splitting on $\mathcal{I}_2$ such that $\mathring{\bbeta}(\Omega)$ and $\mathring{B}_t^{(jo)}$ are computed based on disjoint folds of samples. For a more efficient use of samples, we avoid another sample splitting in the proposed method.

In the following, we present a counterpart of Theorem~\ref{thm-dg} with respect to the max norm. Let $\bar{\lambda}_{\Omega}=\max_{1\leq t\leq q}\|\bM_t[\bbeta(\Omega)]\|_2$.
\begin{theorem}[Domain generalization bounds in max norm]
\label{thm-dg2}
Assume Conditions \ref{cond1}, \ref{cond2}, and \ref{cond-eigen} hold. Assume that $p(\log|\mathcal{O}|+\log n)\leq c_1n$, $\omega_t\leq c_1 |\mathcal{C}_t|n$, and $r_t\leq c_1na_t$, $t=1,\dots,q$ for some small enough constant $c_1$. 
In addition, if $\bar{\lambda}_{\Omega}\geq \sqrt{\frac{\prod_{t=0}^qr_t+\log n}{n}}$ and  $\bar{\lambda}_{\Omega}C_R^{q}\sum_{t=1}^q\sqrt{\frac{r_t+\log n}{na_te_*}}\leq \max_{g\in\mathcal{G}}\min_{t\leq q} \bar{\lambda}_{g_{-t}\circ_t\Omega_t}$, then with probability at least $1-\exp\{-c_1\log n\}$,
\begin{align}
& \max_{g\in\mathcal{G}}\|\hat{\bbeta}^{(g)}-\bbeta^{(g)}\|_2\vee \max_{g\in\mathcal{G}}\frac{1}{\sqrt{n^{(g)}}}\|X^{(g)}(\hat{\bbeta}^{(g)}-\bbeta^{(g)})\|_2\lesssim  \max_{g\in\mathcal{G}}\sum_{t=1}^q\bar{\lambda}_{g_{-t}\circ_t\Omega_t}\sqrt{\frac{C_R^2r_t+\log p_t+\log n}{a_tne_*}}\nonumber\\
&\quad+\max_{g\in\mathcal{G}}\|\bbeta^{(g)}\|_2\sqrt{\frac{p+\log n}{|\mathcal{O}|ne_*}}+C_R^q\sqrt{\frac{r_0+\log|\mathcal{G}|+\log n}{n}}.\label{eq1-max}
\end{align}
\end{theorem}
The results in max norm (\ref{eq1-max}) can be more useful in the domain generalization setting. We see that the rate in max norm is no slower than the rate in vectorized $\ell_2$-norm. The terms $\log p_t$ and $\log|\mathcal{G}|$ appear in the upper bound as we take maximum over all the groups in $\mathcal{G}$.

\subsection{Minimax lower bound}
\label{sec4-2}
In this section, we provide minimax lower bound results for the current problem. Let $\bbeta(\mathcal{G})\in\R^{p\times p_1\times\dots\times p_q}$ be the coefficient tensor corresponding to  a set of group $\mathcal{G}$. 
We consider the following parameter space
\begin{align*}
\Theta(\bm{r};\underline{\lambda},\underline{\bm{\Lambda}},e^*)&=\left\{\bbeta(\mathcal{G})\in\R^{p\times p_1\times\dots\times p_q}: \textup{rank}(\bM_k[\bbeta(\mathcal{G})])=\textup{rank}(\bM_k[\bbeta(\Omega)])\leq r_k, k=0,\dots,q, \right.\\
&\quad\quad \left. \min_{0\leq k\leq q}\Lambda_{r_k}(\bM_k[\bbeta(\mathcal{G})])\geq \underline{\lambda}, ~\Lambda_{r_t}(\bM_t[\bbeta(\mathcal{G}_{-t}\circ_t\Omega_t)])\geq \underline{\lambda}_{\mathcal{G}_{-t}\circ_t\Omega_t},\right.\\
&\quad\quad\left. \max_{0\leq t\leq q}\Lambda_{\max}((B_t^{(jo)})^\top B_t^{(jo)}/a_t)\leq e^*, ~\mathcal{O}\supseteq\Omega\cup_{t=1}^q (\calA_t\circ_t [p_t])\right\},
\end{align*}
where $\bm{r}=(r_0,\dots,r_q)$ and $\underline{\bm{\Lambda}}=\{\underline{\lambda}_{\mathcal{G}_{-t}\circ_t\Omega_t}\}_{t=1}^q$, and $B_t^{(jo)}=\bM_t[\bbeta(\calA_t\circ_t\Omega_t)]\in\R^{\omega_t\times a_tp}$. We present the minimax lower bound result below. 
\begin{theorem}[Minimax lower bound in Frobenius norm]
\label{thm-lb}
Assume Conditions \ref{cond1} and \ref{cond2}, $p_t\geq 3r_t$ for $t=0,\dots,q$, and $q$ is finite. There exists some positive constant $c_1$ such that
\begin{align*}
\inf_{\widehat{\bbeta}(\mathcal{G})}\sup_{\bbeta(\mathcal{G})\in\Theta(\bm{r};\underline{\lambda},\underline{\bm{\Lambda}},e^*)}\P\left(\|\hat{\bbeta}(\mathcal{G})-\bbeta(\mathcal{G})\|_{\ell_2}\geq c_1\sum_{t=1}^q\underline{\lambda}_{\mathcal{G}_{-t}\circ_t\Omega_t}\sqrt{\frac{p_tr_t}{a_tne^*}}+c_1\underline{\lambda}\sqrt{\frac{pr_0}{e^*|\mathcal{O}|n}}+c_1\sqrt{\frac{\prod_{t=0}^qr_t}{n}}\right)\geq 1/4.
\end{align*}
\end{theorem}
Compared with the rate in \eqref{eq1-frob} of Theorem~\ref{thm-dg}, we see that the proposed algorithm is minimax rate optimal in the parameter space $\Theta(\bm{r};\underline{\lambda},\underline{\bm{\Lambda}},e^*)$  given that $e_*\asymp e^*, \underline{\lambda}_{\mathcal{G}_{-t}\circ_t\Omega_t}\asymp\bar{\lambda}_{\mathcal{G}_{-t}\circ_t\Omega_t}$, and $\bar{C}_R=O(1)$.

To summarize, the proposed TensorDG method enjoys both computational efficiency and robustness. If the low-rank model holds, then it has fast convergence rates for unseen domains. On the other hand, 
our proposal could fail if Condition \ref{cond1} is not satisfied. In practice, Condition \ref{cond1} can be diagnosed to some extent. For instance, one can determine whether the rank of $\bM_t[\bbeta(\calA_t\circ_t \Omega_t)]$ equals the rank of $\bM_t[\bbeta(\calA_t\circ_t [p_t])]$ based on the information criterion or the eigen-ratio criterion \citep{han2022rank}. If they are not equal, then Condition \ref{cond1} is violated. 
Such tests can answer the important question whether the model is generalizable based on the observed data, which concerns the reliability of a domain generalization method. In contrast, there seems no direct way to verify the generalizability of invariant causal models.

\section{Extensions}
\label{sec-ext}
In this section, we extend the main methodology to handle transfer learning tasks and high-dimensional domain generalization where the number of covariates can be much larger than the sample size.

\subsection{Extension to transfer learning tasks}
Our method has demonstrated the capability to estimate $\bbeta^{(g)}$ even if $n^{(g)}=0$ by leveraging the low-rank tensor structure. In real-world scenarios, it is common to have only a limited number of samples available from the target domain. Specifically, for a given target domain $g^*$, it often holds that $0<n^{(g^*)}\ll \sum_{g\in\mathcal{O}}n^{(g)}$.  When faced with this situation, we would like to harness the $n^{(g^*)}$ samples from the target domain. \citet{li2022transfer}, \citet{tian2022transfer} , and \citet{li2023estimation} have studied transfer learning with sparsity-based similarity characterizations among many others. Besides, as discussed in Section \ref{sec2-1},  \citet{tripuraneni2021provable} consider transfer learning with low-rankness-based similarity characterizations.

In transfer learning, avoiding negative transfer is a critical concern. Negative transfer occurs when the assumed similarity between the source and target tasks is not upheld, which could lead to a deterioration in the performance of the proposed transfer learning method compared to only using target data. %In transfer learning, a crucial topic is to avoid negative transfer. That is, the similarity assumption on the source and target tasks can be violated and hence the proposed transfer learning method can be worse than only using the target data.  
In our case, the low-rank tensor model---expressed by equation (\ref{model2}) and Condition \ref{cond1}---may not hold for a target domain  $g^*$. 
%  In our case, the low-rank tensor model (equation (\ref{model2}) and Condition \ref{cond1}) can be misspecified for a target domain $g^*$. 
Therefore, we relax (\ref{model2}) to account for an additional level of model heterogeneity. Specifically, for the target model $g^*$ we allow
\begin{align}
\label{model3}
\E[y_i^{(g^*)}|\bx_i^{(g^*)}]=\bx_i^{(g^*)}\bm\gam^{(g^*)},~~\bm\gam^{(g^*)}=\bbeta^{(g^*)}+\bm\delta^{(g^*)},
\end{align}
where $\bbeta^{(g^*)}$ belongs to the tensor $\bbeta(\mathcal{G})$ satisfying Condition \ref{cond1} and $\bm\delta^{(g^*)}\in\R^p$ represents a unique direction of $\bbeta^{(g^*)}$. The magnitude of $\bbeta^{(g^*)}$ also denotes the level of misspecification of the low-rank tensor model. Let $\hat{\bbeta}^{(g^*)}=\{\widehat{\bbeta}(\mathcal{G})\}_{.,i_1,\dots,i_q}$ such that $g^*=(i_1,\dots,i_q)$.
To estimate $\bm\gam^{(g^*)}$, we use the TensorDG estimate $\hat{\bbeta}^{(g^*)}$ as a summary of all the source tasks and perform bias-correction as in oracle Trans-Lasso \citep{li2022transfer}.

\begin{algorithm}[H]
\textbf{Input}: TensorDG estimate $\hat{\bbeta}^{(g^*)}$ and target samples $(X^{(g^*)},y^{(g^*)})$.

\textbf{Output}: $\hat{\bm\gamma}^{(g^*)}$.

\textbf{Step 1:} Compute 
\begin{align*}
\hat{\bm{\delta}}^{(g^*)}=\argmin_{\bm\delta\in\R^p}\left\{\frac{1}{n^{(g^*)}}\|\bm{y}^{(g^*)}-X^{(g^*)}\hat{\bbeta}^{(g^*)}-X^{(g^*)}\bm\delta\|_2^2+\lambda_{g^*}\|\bm\delta\|_1\right\},
\end{align*}
where $\lambda_{g^*}$ is a tuning parameter.

\textbf{Step 2:} Output $\hat{\bm{\gam}}^{(g^*)}=\hat{\bbeta}^{(g^*)}+\hat{\bm{\delta}}^{(g^*)}$.
\caption{TensorTL: Transfer learning based on Algorithm \ref{alg-dg}}
\label{alg-tl}
\end{algorithm}
%In Algorithm \ref{alg-tl}, target samples are partitioned into two folds. The data in $I_2^{(g^*)}$ serve as test samples, aiding in learning the optimal weights for two candidate estimators. Given that $\omega$ is a scalar, the test sample size can be relatively small and hence in Step 0, we only require $|I_2^{(g^*)}|\geq c_0n^{(g^*)}$ for some positive constant $c_0$. The aggregation step (Step 2) has been similarly considered in \citet{li2022transfer} for positive transfer warrantee. More sophisticated aggregation estimators were studied in \citet{rigollet2011exponential} and \citet{dai2012deviation}, which have optimal rates and constants.

\begin{theorem}[Estimation and prediciton errors of TensorTL]
\label{thm-tl}
Assume the Conditions of Theorem \ref{thm-dg2} and model (\ref{model3}). For $\lambda_{g^*}\geq c_0\sqrt{\log p/n^{(g^*)}}$ with a large enough constant $c_0$, it holds that with probability at least $1-\exp\{-c_1\log p\}-\exp\{-c_2\log n\}$,
\begin{align*}
&\|\hat{\bm\gam}^{(g^*)}-\bm\gam^{(g^*)}\|_2\vee \frac{1}{\sqrt{n^{(g^*)}}}\|X^{(g^*)}(\hat{\bm\gam}^{(g^*)}-\bm\gam^{(g^*)})\|_2\lesssim\sum_{t=1}^q\bar{\lambda}_{g^*_{-t}\circ_t\Omega_t}\sqrt{\frac{C_R^2r_t+\log n}{a_tne_*}}\nonumber\\
&\quad+\|\bbeta^{(g^*)}\|_2\sqrt{\frac{p+\log n}{|\mathcal{O}|ne_*}}+C_R^q\sqrt{\frac{r_0+\log n}{n}}+\sqrt{\frac{\|\bm\delta^{(g^*)}\|_0\log p}{n^{(g^*)}}}.
\end{align*}
\end{theorem}
In Theorem \ref{thm-tl}, we establish the convergence rate of Algorithm \ref{alg-tl}. The first three terms in the right-hand-side is the upper bound for the TensorDG estimate $\|\hat{\bbeta}^{(g^*)}-\bbeta^{(g^*)}\|_2$. Recall that the single-task OLS estimator has a convergence rate of order $p/n^{(g^*)}$. Under the mild conditions that $C_R=O(1)$ and $r_0+\log n+\|\bm\delta^{(g^*)}\|_0\log p\ll p$, the TensorTL estimate $\hat{\bm\gam}^{(g^*)}$ has a faster convergence rate than the OLS. Indeed, this conditions requires that the misspecified parameter $\bm\delta^{(g^*)}$ is sparse, i.e., the misspecification level is relatively low.

\subsection{Extensions to high-dimensional scenarios}
In this subsection, we extend the proposed algorithm to the high-dimensional setting where $p$ can be possibly larger than $n$. In this case, the OLS estimate based on each group is no longer feasible. Following the high-dimensional statistics literature, we consider sparse models in this high-dimensional setting.

Given that $\bbeta^{(g)}$ exhibits a sparse pattern, $\ell_1$-penalized methods like the Lasso \citep{tibshirani1996regression} can be employed to replace OLS within Algorithm \ref{alg-dg}. However, a challenge arises since Lasso estimates are inherently biased, potentially resulting in significant estimation errors. Given that $\bM_0[\bbeta(\mathcal{O})]$ is both low-rank and column-wise sparse, it is plausible to assume that $\bbeta^{(g)}$  has a similar support across each $g\in\mathcal{O}$. As such, the support of $\bbeta^{(g)}$   can be estimated utilizing a group Lasso penalty, represented as
%Suppose $\beta^{(g)}$ has a sparse pattern, then $\ell_1$-penalized methods such as the Lasso can be applied to replace the OLS estimates in Algorithm \ref{alg-dg}.  However, the Lasso estimates are biased which may lead to large estimation errors.  As $\bM_0[\bbeta(\mathcal{O})]$ is low-rank and is column-wise sparse, it is reasonable to assume that $\bbeta^{(g)}$ has similar support for each $g\in\mathcal{O}$. Hence, one can estimate the support of $\bbeta^{(g)}$ via a group Lasso penalty. Specifically,
\begin{align}
\label{grp-lasso}
\{\hat{\bbeta}^{(g)}\}_{g\in\mathcal{O}}=\argmin_{\bm{b}^{(g)}\in\R^p, g\in\mathcal{O}}\frac{1}{\sum_{g\in\mathcal{O}}n^{(g)}}\sum_{g\in\mathcal{O}}\|\by^{(g)}-X^{(g)}\bm{b}^{(g)}\|_2^2+\lambda\sum_{j=1}^p \sqrt{\sum_{g\in\mathcal{O}}(b^{(g)}_j)^2},
\end{align}
where $\lambda>0$ is the tuning parameter. Then we can estimate the support via
\begin{align}
\label{eq-Shat}
   \widehat{S}=\left\{1\leq j\leq p:\sqrt{\sum_{g\in\mathcal{O}}(\hat{\beta}^{(g)}_j)^2}\geq \lambda\right\}.
\end{align}
The selection consistency of group Lasso has been studied in \citet{nardi2008asymptotic} under group irrepresentable conditions and in \citet{wei2010consistent} under the sparse Riesz condition.
If $S=\cup_{g\in\mathcal{O}}supp(\bbeta^{(g)})$ can be consistently estimated and $|S|\leq c_1n$ for some small enough constant $c_1$, one can leverage post-selection least square estimates as the starting point. Specifically, we only need to replace $X^{(g)}$ with $X_{.,\widehat{S}}^{(g)}$ in Algorithm \ref{alg-dg} and it produces the estimated tensor $\bbeta(\mathcal{G})_{\widehat{S},.}\in\R^{|\widehat{S}|\times p_1\times\dots\times p_q}$. Due to the sparsity assumption, we can estimate $\bbeta(\mathcal{G})_{\widehat{S}^c,.}=\bf{0}$.

\section{Numerical Experiments}
\label{sec-sim}

We evaluate the performance of our proposals in multiple numerical experiments in comparison to some existing methods. The code for all the methods is available at \url{https://github.com/saili0103/TensorDG}.
\subsection{Domain generalization performance}
\label{sec-sim-1}
We evaluate the dependence of domain generalization errors on $|\calA_t|$, $|\Omega_t|$ and $r_t$.
For a generic estimator $\hat{\bbeta}(\mathcal{G})$, define its Average $\ell_2$-Error (AL2E) as $\|\widehat{\bbeta}(\mathcal{G})-\bbeta(\mathcal{G})\|_{\ell_2}/\sqrt{|\mathcal{G}|}$ and its
Average Domain Generalization Errors (ADGE) as
$\textup{ADGE}=\sqrt{\sum_{g\in\mathcal{O}^c}\|\hat{\bbeta}^{(g)}-\bbeta^{(g)}\|_2^2/|\mathcal{O}^c|}$.
%For each method, we will evaluate $\textup{AGSE}(\mathcal{O})$ and $\textup{AGSE}(\mathcal{O}^c)$ which are the average group-wise errors for the training and test groups. $\textup{AGSE}(\mathcal{O})$ is also known as in-distribution estimation error and $\textup{AGSE}(\mathcal{O}^c)$ is known as out-of-distribution estimation error.

We compare the performance of TensorDG, single-task OLS, and Maximin estimator. It is known that sample splitting can result in inefficient use of samples. We evaluate different versions of sample splitting and find that the most effective version is to all the samples in all the steps of Algorithm \ref{alg-dg}. %Details of the algorithm are included in Section \ref{sec-ap1} in the supplements. 
To compute single-task OLS, we generate $n^{(g)}$ samples for group $g$ if $g\notin\mathcal{O}$. %The single-task OLS is computed based on all the $n^{(g)}$ samples in domain $g$.
In contrast, TensorDG only uses samples in $\mathcal{O}$.

\begin{figure}[H]
\includegraphics[width=0.99\textwidth]{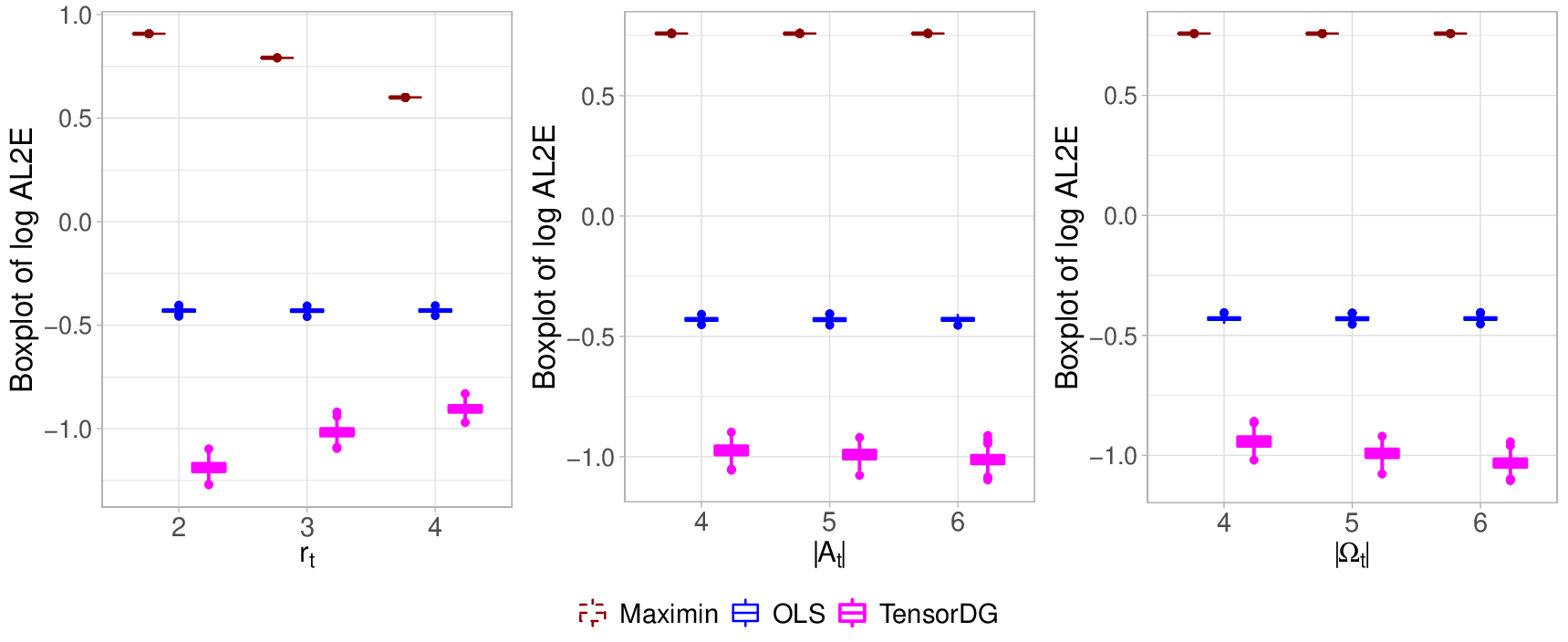}
\caption{Boxplot of AL2E in $\log_5$-scale based on Maximin(dashed brown), OLS(solid blue), and TensorDG(bold solid magenta) in experiments (a), (b), and (c). Each setting is replicated with 500 Monte Carlo experiments.} 
\label{fig0-simu}
\end{figure}

\begin{figure}[H]
\includegraphics[width=0.99\textwidth]{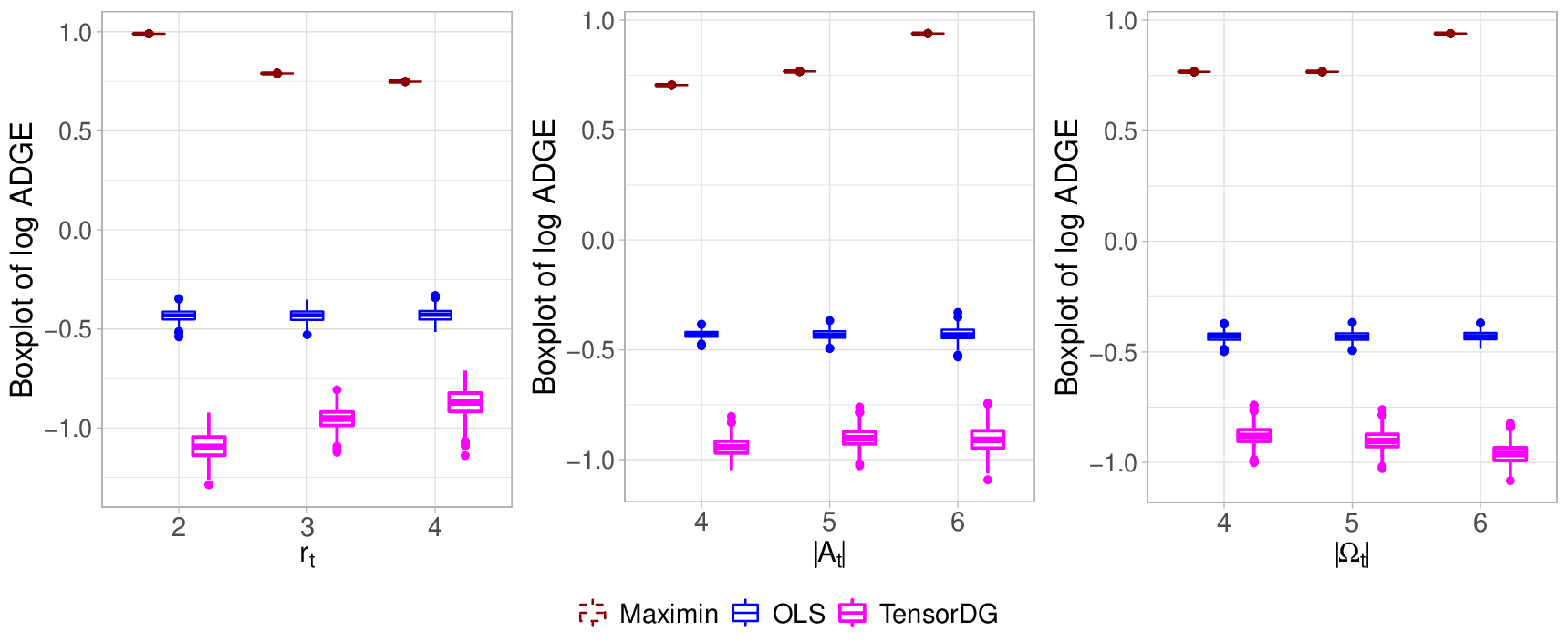}
\caption{Boxplot of ADGE in $\log_5$-scale based on Maximin(dashed brown), OLS(solid blue), and TensorDG(bold solid magenta) in experiments (a), (b), and (c). Each setting is replicated with 500 Monte Carlo experiments.} 
\label{fig1-simu}
\end{figure}

The default setting in our simulation is $n^{(g)}=300$ for each $g\in\mathcal{O}$ , $q=2$, $(p_0,p_1,p_2)=(60,8,8)$, $r_t=3$, $r_0=2r_t$, and $|\calA_t|=|\Omega_t|=5$ for $t=1,2$.  In experiment (a), we consider $r_t\in\{2,3,4\}$ and set other parameters as default. In experiment (b), we consider $|\calA_t|\in\{4,5,6\}$ and set other parameters as default. In experiment (c), we consider $|\Omega_t|\in\{4,5,6\}$ and set other parameters as default. 
The average $\ell_2$-errors of TensorDG, single-task OLS, and Maximin estimator are given in Figure \ref{fig0-simu}. We see that the average estimation error of TensorDG increases as $r_t$ increases and decreases as $|\calA_t|$ or $|\Omega_t|$ increases, which aligns with our theoretical analysis. In Figure \ref{fig1-simu}, we see that TensorDG has the smallest average domain generalization errors. 
The single-task OLS has larger errors as it only uses $n$ samples from a single group and its estimation accuracy is invariant to $r_t$, $|\calA_t|$, and $|\Omega_t|$. The Maximin estimator has the largest domain generalization errors. One reason is that the reward function that Maximin estimator maximizes, e.g., (7) in \citet{meinshausen2015maximin} and (6) in \citet{guo2023statistical}, is different from prediction errors or estimation errors under general conditions.

\subsection{Transfer learning performance}
\label{sec-sim-2}

Next, we evaluate the performance of TensorTL for transfer learning tasks. For comparison, we consider a modification of Meta-LM-MoM proposed in \citet{tripuraneni2021provable}, where ``MoM'' refers to the method-of-moments for estimating the linear subspace of $\bM_0[\bbeta(\mathcal{G})]$. We modify the original Meta-LM-MoM by changing the MoM step to our proposed Algorithm \ref{alg-svd} for mode 0 because we find that our proposal can significantly improve the estimation accuracy over MoM. We call this transfer learning method ``Meta-LM*''. We consider settings same as the ones in Figure \ref{fig1-simu} except that $n^{(g)}=150$ for $g\in \mathcal{O}^c$. This setup is due to in transfer learning settings, the data from the target domain is always very limited. For $\bm\gam^{(g^*)}$ defined in (\ref{model3}), we consider $\|\bm\delta^{(g^*)}\|_0\in\{0,3\}$, respectively.  If $\delta^{(g^*)}_j\neq 0$, we simulate $\delta^{(g^*)}_j\sim N(0,0.25)$ independently.  We report the boxplot of its Transfer Learning Error (TLE)
$\|\hat{\bm{b}}^{(g^*)}-\bm\gamma^{(g^*)}\|_2$ for all $g^*\in\mathcal{O}^c$, where $\hat{\bm{b}}^{(g^*)}$ denotes a generic transfer learning estimate of $\bm\gam^{(g^*)}$.

\begin{figure}[H]
\includegraphics[width=0.99\textwidth]{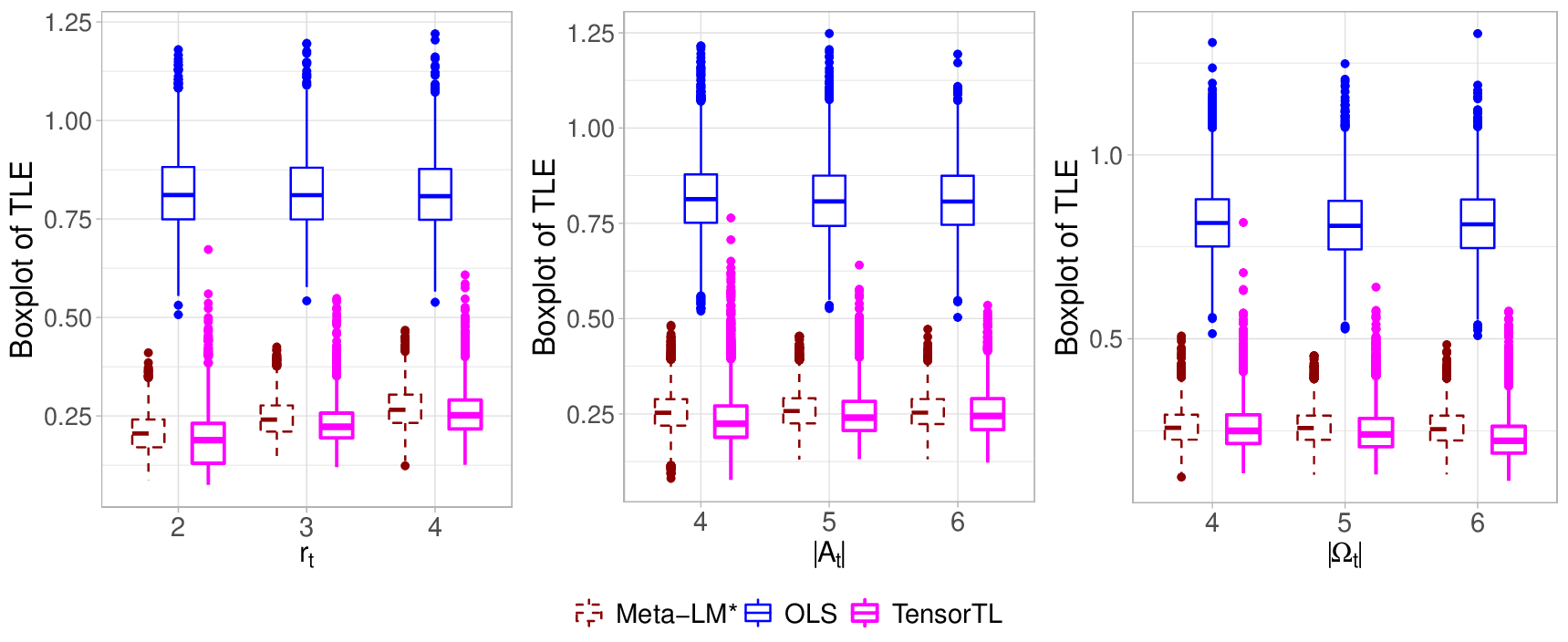}
\caption{Boxplots of TLE based on Meta-LM*(dashed brown), OLS(solid blue), and TensorTL (bold solid magenta)  in experiments (a), (b), and (c) with $\|\bm\delta^{(g^*)}\|_0=0$. Each setting is replicated with 500 Monte Carlo experiments.}
\label{fig2-simu}
\end{figure}

\begin{figure}[H]
\includegraphics[width=0.99\textwidth]{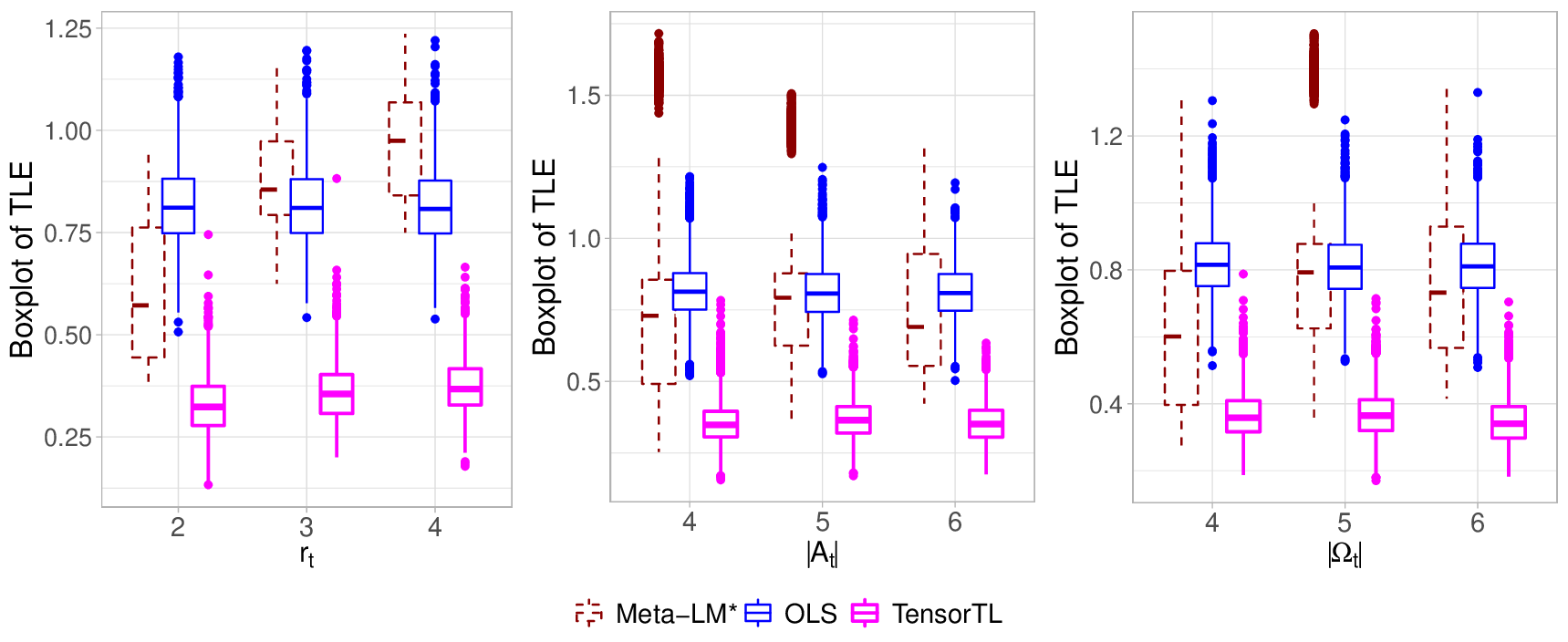}
\caption{Boxplots of TLE based on Meta-LM*(dashed brown), OLS(solid blue), and TensorTL (bold solid magenta)  in experiments (a), (b), and (c) with $\|\bm\delta^{(g^*)}\|_0=3$. Each setting is replicated with 500 Monte Carlo experiments.}
\label{fig3-simu}
\end{figure}

From Figure \ref{fig2-simu}, we see that Meta-LM* and TensorTL improve over single-task OLS when the low-rank tensor model is correctly specified. Meta-LM* has slightly larger estimation errors than TensorTL in most settings. This is because its accuracy relies on $n^{(g^*)}$ which is relatively small in these experiments.  In contrast, the performance of TensorTL is comparable to TensorDG if the low-rank tensor model is correctly specified and hence its performance is not limited by $n^{(g^*)}$. In Figure \ref{fig3-simu}, we consider the case where the low-rank tensor model is mis-specified. We see that Meta-LM* can be even worse than OLS in this case but TensorTL is still robust. This demonstrates the effectiveness of the bias-correction step in TensorTL.

%In the high-dimensional setting, we set $p=400$ and all other dimensions are the same as in Section \ref{sec-sim-1}. We set $\beta^{(g)}_j=0$ for $50<j\leq p$. We perform (\ref{grp-lasso}) via R package ``gglasso'' \citep{gglasso} to first select $\widehat{S}$ via (\ref{eq-Shat}) with $\lambda=0.1\sqrt{\log p/n}$.

\section{Real Data Application}
\label{sec-data}
We apply the proposed methods to the ``Adult'' dataset \citep{kohavi-nbtree} to predict the education achievements for different ethnic groups. The response variable takes integer values from 1 to 16 indicating education levels from preschool to doctorate degree. We consider two specifications of group indices. 

\begin{table}[!htbp]
\centering
\begin{tabular}{|c|c|c|c|c|c|}
\hline
&White &Black &API &  AIE & Other\\
\hline
M & 19174 & 1569 & 693 &192 & 162\\
F & 8642 & 1555& 346 & 119 & 109\\
\hline
\end{tabular}
\caption{Unweighted sample size for 10 groups, where ``M'' and ``F'' are short for ``Male'' and ``Female'', respectively, ``API'' is short for ``Asian-Pacific-Islander'', and ``AIE'' is short for ``American-Indian-Eskimo''.}
\label{tab-samplesize}
\end{table}

The first one defines groups by race and gender. 
The sample size of each group is listed in Table \ref{tab-samplesize}. We see that the sample size of ``White Male'' group is about 200 times larger than the sample size of ``American-Indian-Eskimo Female'' group. Therefore, domain generalization and transfer learning methods can be helpful in this case. We set test domains as two domains with the smallest sample sizes, i.e., $\mathcal{O}^c=\{\text{AIE-F}, \text{Other-F}\}$. We run TensorDG based on data in $\mathcal{O}$ and further perform Algorithm \ref{alg-tl} for TensorTL based on training data for the each target domain. 
After converting categorical variables to dummy variables, we arrive at 79 covariates which exhibit colinearity. Therefore, for TensorDG and TensorTL, we first apply Lasso to remove variables that are not predictive in all the observed groups.
We also compute Lasso as the benchmark single-task method. For comparison, we compute the Maximin estimator based on the Lasso, which only uses data in $\mathcal{O}$ and compute ``Meta-LM*'' based on the variables selected by the Lasso, which is a transfer learning method as described in the simulation section. This survey data set also has a weight variable for each observation and we take the weights into consideration in the training and testing phases.

%\begin{figure}[H]
%\centering
%\includegraphics[height=7.3cm,width=0.49\textwidth]{Adult-Tensor-OLS-Lasso}
%\includegraphics[height=7.3cm,width=0.49\textwidth]{Adult-Tensor-MM-Lasso}
%\caption{Group-wise classification errors (GCE) for each target domain with TensorTL, single-task sparse LDA, and the Maximin estimator for different ethnic groups based on sparse LDA models. The red line is the 45-degree angle line.}
%\label{fig-data1}
%\end{figure}
\begin{figure}[H]
\centering
\includegraphics[height=5.5cm,width=6cm]{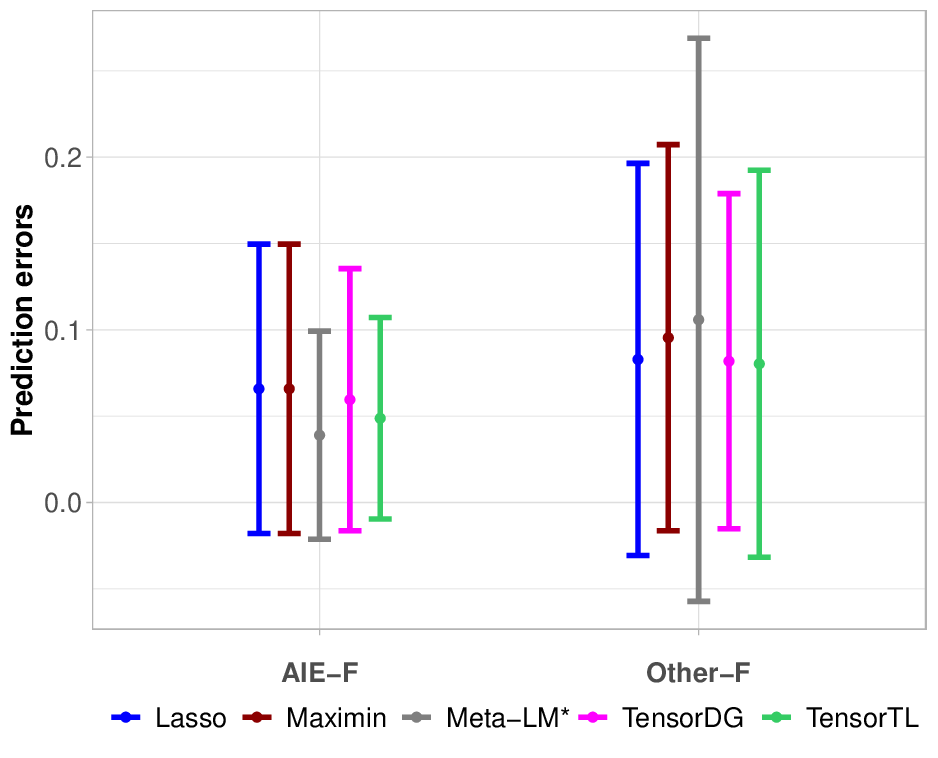}
\includegraphics[height=5.5cm,width=6cm]{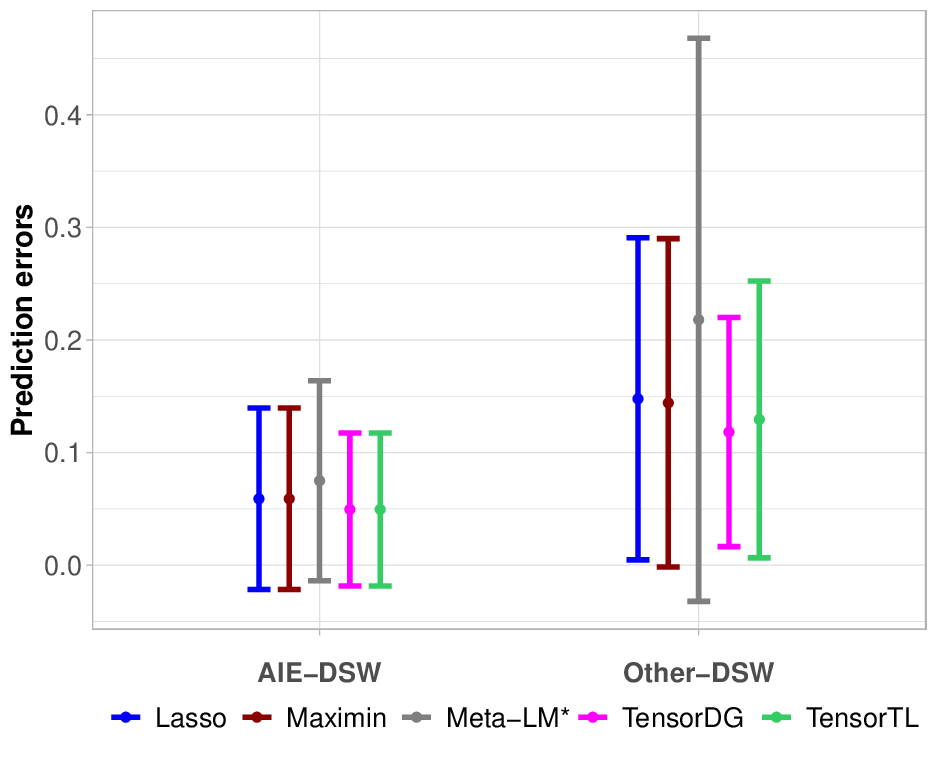}
\caption{Left panel: prediction errors for two minority groups with groups defined by race and gender based on different methods. Right panel: prediction errors for two minority groups with groups defined by race and marital status based on different methods. The prediction errors for the $i$-th observation are scaled by $y_i^2$. The center and width of each error bar denote the mean and the standard error, respectively.}
\label{fig-data1}
\end{figure}

In the second experiment, we define groups by race and marital status. Specifically, we consider three marital status ``Never married'', ``Married'', and ``Divorced, separated, or widowed (DSW)''. This gives 15 groups in total and the group sizes are given in Table \ref{tab-samplesize2}. In this experiment, we set test domains as two domains with the smallest sample sizes, i.e., $\mathcal{O}^c=\{\text{AIE-DSW}, \text{Other-DSW}\}$. We also evaluate the prediction errors of the five methods described above. 
\begin{table}[!htbp]
\centering
\begin{tabular}{|c|c|c|c|c|c|}
\hline
&White &Black &API &  AIE & Other\\
\hline
Never married & 8757 & 1346 & 372 & 103 & 105\\
Married & 13723 & 900& 549 & 125& 120\\
DSW & 5336& 878&118&83 &46\\
\hline
\end{tabular}
\caption{Unweighted sample size for 15 groups, where ``DSW'' is short for ``Divorced, separated, or widowed'', ``API'' is short for ``Asian-Pacific-Islander'', and ``AIE'' is short for ``Amer-Indian-Eskimo''.}
\label{tab-samplesize2}
\end{table}

The results for these two experiments are reported in Figure \ref{fig-data1}. TensorDG and TensorTL have better prediction accuracy than the single-task Lasso in all the test domains. For group ``AIE-F'', TensorDG has slightly larger errors than TensorTL, which implies that model (\ref{model3}) is more suitable than the low-rank tensor model for this target group. 
Maximin and Meta-LM* have large standard errors and are no better than the single-task Lasso in three out of four experiments. 

\section{Discussion}\label{sec:discussion}
We study domain generalization and transfer learning with multi-dimensional group indices in linear models. Based on a low-rank tensor model, we develop rate optimal methods for domain generalization and demonstrate its reliable performance in numerical studies. The proposed framework can be extended to other settings. %While we focus on structural missing, i.e., the observed groups contain a body set and arm sets, it is possible that the observed groups are randomly selected from all the groups. In this case, tensor completion methods for uniformly random missing patterns can be derived in our framework for domain generalization and transfer learning. 
An important extension is to deal with binary or categorical outcomes based on other machine learning methods. As deep neural networks have shown significant successes in practice, a direction of interest is to extend the current model to deep neural nets where each layer of neural networks across different domains forms a low-rank tensor. The technical tools developed in this paper can potentially apply to these cases to facilitate developing domain generalization in neural networks and other machine learning methods with provable guarantees. 

\section*{Acknowledgement}
Sai Li's research was supported in part by the National Natural Science Foundation of China (grant no. 12201630). Linjun Zhang's research was supported in part by NSF grant DMS-2015378.

\bibliographystyle{chicago}
\bibliography{dg-tensor}
\end{document}